\documentclass[review,a4paper, 11pt]{elsarticle}
\usepackage{hyperref}
\usepackage{amsmath,amssymb,amsfonts,bbm, amsthm}
\usepackage{algorithm,algorithmic}
\usepackage{graphicx}
\usepackage{url}
\usepackage{booktabs} 
\usepackage{textcomp}
\usepackage{xcolor,color, colortbl}
\usepackage{multirow}
\usepackage{natbib}
\usepackage[left=1in,right=1in,top=1in, bottom=1in]{geometry}
\definecolor{Gray}{gray}{0.9}

\newcommand{\slfrac}[2]{\left.#1\middle/#2\right.}
\journal{Signal Processing}
\makeatletter
\def\ps@pprintTitle{%
  \let\@oddhead\@empty
  \let\@evenhead\@empty
  \let\@oddfoot\@empty
  \let\@evenfoot\@oddfoot
}
\makeatother
\newtheorem{theorem}{Theorem}
\newtheorem{fact}{Fact}
\DeclareMathOperator*{\argmin}{arg\,min}
\def\env{\mbox{\footnotesize env}}
\def\app{\mbox{\footnotesize appopt}}
\def\subopt{\mbox{\footnotesize subopt}}
\def\opt{\mbox{\footnotesize opt}}
\def\proj{\mbox{\footnotesize proj}}
\def\eE{{\mathbb E}}
\def\pP{{\mathbb P}}
\def\zZ{{\mathbb Z}}
\def\SA{\mbox{SA}}
\graphicspath{{Figures/}} 

\begin{document}

\begin{frontmatter}

\title{Overpredictive Signal Analytics in Federated Learning: \\ Algorithms and Analysis\tnoteref{mytitlenote}}

	\author{Vijay Anavangot}
\address{Department of Electrical Engineering, \\ Indian Institute of Technology Bombay, \\ Mumbai - 400076, India \\
	Email: a.vijay.2014@ieee.org}
%
%
%
%
\begin{abstract}
	%
	Edge signal processing facilitates distributed learning and inference
	in the client-server model proposed in federated learning.
	In traditional machine learning, clients (IoT devices) that acquire raw signal
	samples can aid a data center (server) learn a global signal model by
	pooling these distributed samples at a third-party location.
	Despite the promising capabilities of IoTs, these distributed deployments often
	face the challenge of sensitive private data and communication rate
	constraints.  
	This necessitates a learning approach that communicates a processed
	approximation of the distributed samples instead of the raw signals. 
	Such a decentralized learning approach using signal approximations will be
	termed distributed signal analytics in this work. 
	Overpredictive signal approximations may be desired for distributed signal
	analytics, especially in network demand (capacity) planning
	applications motivated by federated learning. 
	In this work, we propose algorithms that compute an overpredictive signal
	approximation at the client devices using an efficient convex
	optimization framework.  
	Tradeoffs between communication cost, sampling rate, and the signal
	approximation error are quantified using mathematical analysis. 
	We also show the performance of the proposed distributed algorithms on a
	publicly available residential energy consumption dataset.
	%
\end{abstract}
\end{frontmatter}
%
%
\section{Introduction}
%
%
%
Distributed signal (or information) processing using Internet of Things (IoT),
facilitates real time monitoring of signals (such as environmental pollutants,
financial, energy consumption) in a smart city. Despite promising
capabilities, these distributed deployments often face the challenge of data
privacy and communication rate constraints~\cite{Lalitha:2013,McMahan:2017}. 
In traditional machine learning, training data is moved to a data center, 
which requires massive data movement from distributed IoT 
devices to a third-party location, thus raising concerns over privacy and
inefficient use of communication resources. 
Moreover, the growing network size, model size, and data volume combined lead
to unusual complexity in the design of optimization algorithms, beyond the
compute capability of a single device. This necessitates novel system
architectures to ensure stable and secure operations.  
Federated learning (FL) architecture  can be a promising solution for enabling
IoT-based smart city applications, by addressing these
challenges~\cite{Konevcny:2016a, Konevcny:2016b, McMahan:2017,
Bonawitz:2019,Kairouz:2021,Li:2020}.
In the FL paradigm, a global server orchestrates model
training, without raw-data being transferred from the participating devices (or
clients).
Edge-deployed signal processing algorithms, such as sparse
approximations and statistical learning methods will be essential for
efficient management of compute and communication resources in FL.
%
%
%

\begin{figure}[!hbt]
	\centering
	\includegraphics[scale=0.5]{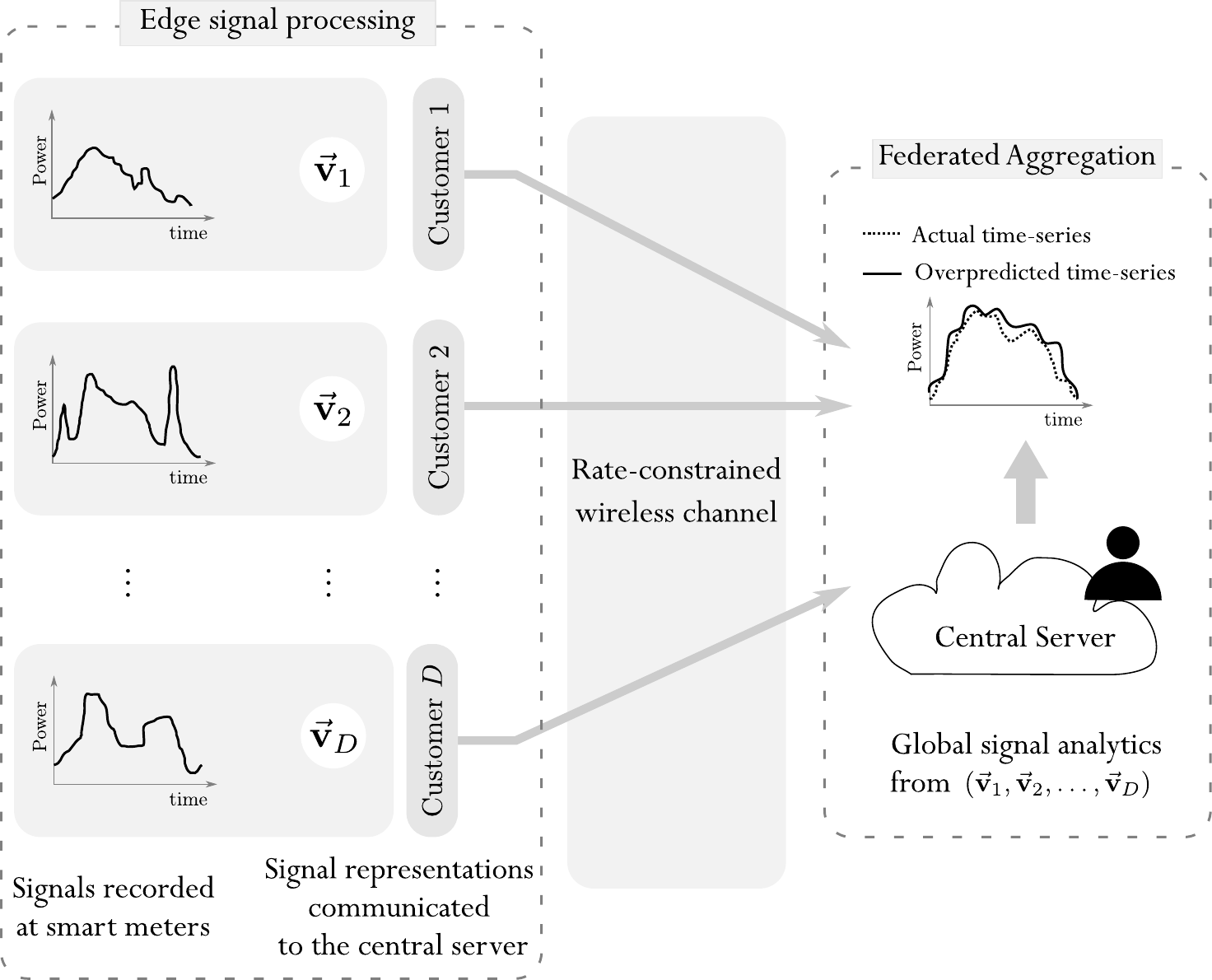}
	\caption{\label{fig:Block_Dia}The figure illustrates federated learning
	for overpredictive signal analytics in a smart city with $D$
	electricity consumers. The power consumption time-series recorded at
	the consumer's smart meters are converted to compressed signal
	representations, $\{\vec{\mathbf{v}}_1,\vec{\mathbf{v}}_2,\ldots,\vec{\mathbf{v}}_D\}$ using edge
	signal processing. These signal representations are communicated to the
	central server for computing signal analytics that overpredict the
	actual demand time-series.} 
\end{figure}

In certain networked system applications, such as electricity network demand
planning, or flood/drought forecasting systems or Unmanned Aerial Vehicle (UAV)
path planning, or autonomous driving it is essential to have an overpredictive
estimate of signals. For instance, consider a smart city application shown in
Fig.~\ref{fig:Block_Dia}, where each household has a smart energy meter to
monitor and record the instantaneous electric power. At the end of the day,
each smart meter will need to report an approximate summary of the consumer
side energy usage to a centrally located server, accessible to the electricity
network planner. The planner desires to have an overpredictive estimate of
the energy demand time-series, so that the generation can be planned to always
meet the consumer demand.
In such a smart city application, we attempt to answer the question,
\textit{how to perform overpredictive signal analytics when the devices are
distributed?}.  The significant challenges in doing such distributed signal
analytics in FL are identified to be: (1) asymmetric communication resources
(i.e. uplink rate is less than the downlink rate), (2) heterogeneous devices,
(3) sensitive private data, and (4) scale of operation (a large number of
devices)~\cite{McMahan:2017}. This work will primarily address communication
rate constraints while implicitly ensuring user privacy and scalability. In
this first exposition, we restrict the focus to the class of homogeneous user
devices.    
FL allows a common global model to be learned from distributed
devices by using efficient compressed signal representations. The existing
literature in FL, however, has a focus on learning a parametric
model using a central parametric server, such as a neural network model
through the successive exchange of gradients~\cite{Konevcny:2016a,
Konevcny:2016b, McMahan:2017, Bonawitz:2019, Suresh:2017, Amiri:2020}.
In this work, the approach is different as we utilize the FL architecture for
learning signal analytics~\cite{AIGoogle:2020}, instead of a
parametric model. The goal is to achieve efficient statistical representation
for performing analytics involving signal overprediction.

A key feature in FL is to utilize edge signal processing to perform distributed
signal analytics. Let us revisit the example of energy demand time-series
aggregation performed at the smart city electricity server, as illustrated in
Fig.~\ref{fig:Block_Dia}. This central server performs network planning to meet
the electricity demands of the consumers in the city. At each of the $D$
consumers (also known as edge/client devices), the installed smart meter would
record the instantaneous power consumed and communicate a compressed
representation, \textit{viz.}
$\{\vec{\mathbf{v}}_1,\vec{\mathbf{v}}_2,\ldots,\vec{\mathbf{v}}_D\}$, of the
demand signal to the server through a rate-constrained wireless channel.  The
network planner aggregates the client signal representations to estimate an
approximate demand time-series signal, which over-predicts the actual aggregate
demand signal. A possible implementation scheme would be to compute approximate
signal representations locally at the edge devices such that these
representations satisfy the individual signal overprediction constraints. The
edge devices then communicate these compressed signal representations to the
server, where signal analytics of interest at computed. The signal analytics
desired at the server may include the peak electric power demand or the
aggregate signal Cumulative Distribution Function (CDF). Such overpredicted
demand signal analytics computed from the approximate signal representations
$\{\vec{\mathbf{v}}_1,\vec{\mathbf{v}}_2,\ldots,\vec{\mathbf{v}}_D\}$, will reduce
the client-server communication cost. Since the actual data does not move to
the server, implicit data privacy is achieved. The computed signal analytics at
the server is useful to maintain the stability and security of the electricity
distribution network.

Our main contributions in this work are summarized below.
%
\begin{itemize}\itemsep1ex
  \item We consider signal approximation based on overprediction constraint in
	  the federated learning setup by using the Fourier basis for signal
		representation.

  \item We develop a aggregation procedure at the central server to learn the
	  global signal analytics, by relying on the empirical CDF of the aggregate
		signal.
  \item We derive mathematical upperbounds on the pointwise difference of the
	  actual signal CDF and the Glivenko-Cantelli CDF estimate of the
		overpredicted signal, considering the effect of signal sampling.
  \item We present the tradeoff between the communication cost and the signal
	  approximation error, and validate it using experiments on a
		publically available residential energy consumption
		dataset~\cite{Mammen:2018}.  
\end{itemize}

\subsection{Prior Literature}
%
Using the federated learning (FL) approach, communication-efficient distributed
learning was introduced by McMahan et al.~\cite{McMahan:2017}. The FL
architecture proposed in this work considers massively distributed,
heterogeneous devices, which have limited communication. The authors illustrate
the utility of FL models for training deep neural networks with on-device data
using stochastic gradient descent (SGD) based \textit{FedAvg} algorithm. An
improved \textit{FedProx} algorithm was later proposed to handle system
heterogeneity in FL, which in addition has convergence
guarantees~\cite{Nilsson:2018}. In another related work, Suresh et
al.~\cite{Suresh:2017} discuss a communication-efficient distributed mean
estimation using FL, stochastic quantization, and structured rotation.
Bonawitz et al.~\cite{Bonawitz:2019}, has described the system-level
implementation of FL algorithms in large-scale networks.  Advances in FL
algorithms, including open directions in distributed learning with privacy
sensitive data, have been summarized in the review paper by Kairouz and
McMahan~\cite{Kairouz:2021}.  From a signal processing perspective, the
applications and challenges of FL have been discussed by Li et
al.~\cite{Li:2020}. Signal processing methods such as compressed gradients and
sketched updates have found utility in training and deploying FL in low-power
TinyML devices~\cite{Alistarh:2017, Feraudo:2020}.  Further, FL models to
address the fairness of resource allocation have been studied from the
perspective of parametrized cost functions and empirical probability
distributions~\cite{Li:2020b, Mohri:2019}. The references above mostly deal
with single task FL models. Considering real-world IoT systems, a multi-task
learning scheme was proposed~\cite{Smith:2017} that allows for a certain degree
of model personalization.

Several applications of signal overprediction are discussed in the literature.
These include watershed management (hydrology), Unmanned Aerial Vehicle (UAV) path
planning, CPU power prediction, and database management for TV whitespace
protection contours. In hydrology, flood and drought predictions are made
possible through statistical learning approaches that use custom loss functions
to include overprediction constraints~\cite{White:2020}. Support Vector Machine
(SVM) models utilizing asymmetric loss functions are generally used for CPU
power cycle prediction~\cite{Stockman:2011}.  Recent research shows that neural
networks do not capture accurate depth information in images, which can have
significant safety implications in autonomous driving~\cite{Dijk:2019}.
Overpredicting ADC design is a suggested solution to overcome this safety
challenge. Maheshwari and Kumar have proposed a novel quantizer design for the
TV whitespace (geo-database management) application, which determines an
overpredicting envelope~\cite{Maheshwari:2015}. In a recent research, a Federated
learning (FL) approach considering personalized client attributes is analyzed
for energy demand prediction using a clustered aggregation
scheme~\cite{Tun:2021}. In another related work, Konstantinos et al. studies
the time-series dimensionality reduction approach with symbolic aggregate
approximation (SAX) and Lloyd's algorithm~\cite{Konstantinos:2021}, using a
combination of quantization and event-based sampling to achieve signal
compression.  Recently, Saputra et al. have examined the utility of FL
algorithms for location-specific demand prediction in an electric vehicle (EV)
charging architecture~\cite{Saputra:2019}. 



\section{Distributed Signal Model and Background Concepts} \label{sec:prob_setup}

In this section we describe the basic signal model using the FL architecture
and discuss the related background concepts for overpredictive signal
analytics. The signal model has been adapted from the abridged paper by
Anavangot and Kumar~\cite{Anavangot:2020}. 

\subsection{Federated Learning Signal Model}

\begin{figure}[!hbt]
  \centering
  \includegraphics[scale=0.8]{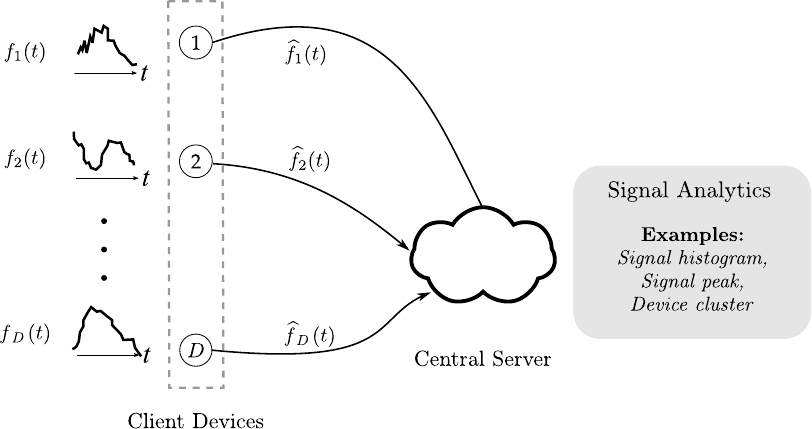}
	\caption{\label{fig:SCApp} The system overview of a federated learning
	model adapted from~\cite{Anavangot:2020}. In this model, the $D$
	participating client devices communicate an approximation of the
	observed signal, and the central server aggregates these signal
	approximations to learn global signal statistics such as histogram,
	signal peak, and device clusters.}
\end{figure}

We consider the federated learning architecture that consists of $D$
distributed client devices and a centrally located server as shown in
Fig.~\ref{fig:SCApp}. Each client device in this model observes a continuous
time signal $f_i(t)$ for $i \in\{1,2,\ldots, D\}$. Without loss of generality
we will assume that each observed signal has a bounded support in the time
interval $[0,1]$. Further, in the ensuing analysis, we also assume that these
signals are $p$-times differentiable, such that $p \geq 1$. Under these signal
assumptions, we devise algorithms to learn a global signal model using
clients-server communication.  Due to rate constrained nature of the
client-server communication channel, the devices may communicate only an
approximate signal, which we denote by $\widehat{f}_i(t)$ for $i \in
\{1,2,\ldots,D\}$.

\subsection{Signal Representation}

In this exposition, we represent the signals using the Fourier basis. That is,
each recorded signal at the clients can be represented by a Fourier series, 
\begin{align*} 
	f_i(t) = \sum_{k = -\infty}^{\infty} a_i[k] \exp(j 2 \pi k t),
	\quad t \in [0,1], 
\end{align*}
where $i \in \{1,2, \ldots,D\}$. As the Fourier series representation is
periodic we will assume $f_i(0) = f_i(1)$. From the $p$-times differentiability
assumption, we observe a polynomial decay of Fourier series coefficients of the
signal, which is stated in the fact below:
\begin{fact}[Sec~2.3,\cite{mallat2008}]
\label{lem:regularity} 
A signal $f(t), t \in [0,1]$, with $f(0) = f(1)$, is $p$-times differentiable
if its Fourier coefficient~$a[k]$ satisfies the condition,
 \begin{align}
|a[k]| \leq \frac{C}{|k|^{p+1+\varepsilon}} \quad \text{ for some }
C, p, \varepsilon >0. \nonumber
\end{align}
\end{fact}
\subsection{Problem Formulation and Related Definitions}

For ths signal approximation, consider the $2L+1$ length bandlimited
approximation of the signals, $f_i(t)$ for $i \in \{1,2,\ldots,D\}$, given by
the Fourier series,
\begin{align} \label{eq:Fourierapp}
  \widehat{f}_i(t) = \sum_{k= -L}^{L} b_i[k] \exp(j 2 \pi kt) \quad \text{ for } t \in
  [0,1], 
\end{align}
where the Fourier coefficients $b_i[k]$ for $k \in \{-L,\ldots,L\}$, will be a
function of $a_i[k], k \in \zZ$. 

The approximation coefficients $b_i[k]$ are computed by solving a constrained
optimization program that minimizes the approximation error with respect to a
distance metric $d(.)$. Typical distance metrics include ${\cal L}_1$, ${\cal L}_2$
or ${\cal L}_\infty$ norm, subject to the overprediction constraint.
The optimization can be stated as,
\begin{align} \label{eq:opt_d_measure}
  \argmin_{\widehat{f}_i(t)} d(f_i(t),\widehat{f}_i(t)) \quad \text{ subject to }
  \widehat{f}_i(t)  \geq f_i(t).
\end{align}

Because the Fourier basis is orthogonal, an equivalent optimization problem can
be framed in terms of the Fourier coefficients $a_i[k]$'s and $b_i[k]$'s.
Therefore, when the distance metric is the ${\cal L}_2$ norm, the equivalent
optimization problem is
\begin{align}
  \argmin_{b_i[k]} \sum_{k = -L}^{L} \left|a_i[k] -
  b_i[k]\right|^2 \;\; \text{ subject to }
  \widehat{f}_i(t)  \geq f_i(t).
\end{align}
The considered problem has linear constraints that meet the overprediction
criteria, hence the signal approximation obtained by this method will be called
the \textit{envelope approximation}. Throughout this paper we will denote the
envelope approximation of a signal $f(t)$ by $\widehat{f}_{\env}(t)$.
\subsection{Choice of the distance metric}
The choice of the distance metrics used to construct the envelope approximation
from the signals recorded at each client device is described.
Note that the envelope approximation $\widehat{f}_{\env}(t)$ of the signal
$f(t)$ is the solution to the optimization problem described in
\eqref{eq:opt_d_measure}, with respect to a distance metric $d(.)$ defined over
the signal space. 
In this paper, we restrict the analysis to the convex distance measures ${\cal L}_1$
and ${\cal L}_2$, and later take up the ${\cal L}_\infty$ formulation to derive
performance bounds.
By using the Fourier representation of $\widehat{f}_{\env}(t)$ (akin
to~\eqref{eq:Fourierapp}, but with the subscript indices dropped) and $f(t)$,
and the envelope property $\widehat{f}_{\env}(t) \geq f(t)$ it can be seen
that,
\begin{align}
  \|\widehat{f}_{\env}-f\|_1 & = b[0] -a[0],\label{eq:L_1} \\[2ex]
  \|\widehat{f}_{\env}- f \|_2^2 = & \sum_{|k| \leq L} |b[k] -a[k]|^2 + \sum_{|k|>L}
	|a[k]|^2 \label{eq:L_2} \\[2ex]
  \|\widehat{f}_{\env}- f\|_\infty \leq & \sum_{|k| \leq L} |b[k] -a[k]| + \sum_{|k|>L}
|a[k]| \label{eq:L_inf} 
\end{align}    
For notational convenience, the signal approximation error corresponding to
${\cal L}_1$, ${\cal L}_2$ and ${\cal L}_\infty$ distance measures will be
denoted by $\SA_1$, $\SA_2$ and $\SA_\infty$ respectively. The upperbound in
\eqref{eq:L_inf} will be used to show the performance bounds of envelope
approximations in Section~\ref{subsec:DC_boost}.


\section{Algorithms for Overpredictive Signal Analytics}
\label{sec:Algo}
In this section, we describe the algorithms to execute envelope
signal analytics using the federated learning model as illustrated in
Fig.~\ref{fig:SCApp}. For the distributed model with $D$ devices, each client
device will report an overpredictive envelope approximation of the signal
observed. These signal approximations are solutions to optimization programs
implemented at the individual client devices. The envelope signals obtained are
later aggregated to compute signal analytics at the central server. The
algorithms presented here is an elaborated version of the the discussions in
Anavangot and Kumar~\cite{Anavangot:2020}.

\subsection{Algorithms at the Clients}
%

The goal of the client devices is to compute the best possible envelope signal
which approximates the true signal. The optimization problem of interest can be
stated as,
\begin{align}
  \min \SA_q := \int_{0}^1 \left| \widehat{f}(t) - f(t) \right|^q \mbox{d}t \mbox{
subject to } \widehat{f}(t) \geq f(t) \label{eq:sa1sa2}
\end{align}
for $q = 1, 2$, where the above minimizations are over the individial
approximate signals $\widehat{f}_1(t), \ldots, \widehat{f}_D(t)$.
These optimization problems can be efficiently solved using the equivalent
Fourier domain representation and the respective norms, described in
\eqref{eq:L_1} and \eqref{eq:L_2}.

We note that the inequality constraints of the problem are linear. Thus, for
$q=1$, the above envelope approximation formulation will become a
\textit{linear program}, since the cost function in the Fourier representation
is the difference of the zero frequency components of the envelope and true
signal.
When $q=2$ the problem is equivalent to a \textit{quadratic program with linear
constraints}, with the objective described by the squared error of the Fourier
coefficients within bandwidth $L$. Both $q=1$ and $q=2$, has linear inequality
constraints due to the envelope overprediction criteria. The choice of the
optimization problem to solve -- ${\cal L}_1$ or ${\cal L}_2$, depends on the
tradeoff between mean squared error, subsampling and envelope violations. As
linear programs have lower computational complexity when compared to quadratic
programs, the client devices with limited hardware can choose to solve ${\cal
L}_1$ optimization problem over the ${\cal L}_2$ problem.
\subsubsection*{Envelope Optimization with Discrete-time Signals}
In practice, the recorded client device signals are available at $n$ discrete time
samples. Since the true signal values are unknown at all times except the
sampled points, the set of inequality constraints in discrete-time case is a finite
subset of the envelope constraints in the continuous time case. Hence, the
discrete-time optimization yields a lower cost than the continuous-time
optimization. A related analysis of the discrete-time envelope optimization in
the context of TV whitespace application, along with Mean Squared Error (MSE)
performance is discussed in~\cite{Kumar:2017}. More detailed analysis of the
discrete time envelope optimization is presented in Sec~\ref{sec:Main_results}.
\subsection{Aggregation at the server}
The signal analytic of interest is computed at the server by combining signal
approximations reported at the edge devices.
A signal analytic, in general, is a function over the envelope approximations
$\widehat{f}_i(t)$ for $i \in \{1,2,\cdots,D\}$. 
In this work we consider two examples of such signal analytics which include
\begin{enumerate}
	\item the aggregate function, 
		\begin{align}
			\widehat{s}(t) := \sum_i \widehat{f}_i(t), \label{eq:AGG}
		\end{align}
	\item the Glivenko-Cantelli empirical CDF, 
\begin{align}
  \widehat{F}_N(x) := \frac1N \sum_{n=1}^{N} \mathbbm{1}_{(-\infty,x]}(\widehat{f}(t_n)), \label{eq:GCC}
\end{align}
where $\mathbbm{1}{(-\infty,x]}(Y)$ represents the $0$--$1$ indicator function for the
probability event $\{Y \leq x\}$, and $t_n$ for $n \in \{1,2,\ldots,N\}$ represents
the time samples in the time interval $[0,1]$.
\end{enumerate}
This classical Glivenko-Cantelli extimate of the CDF~\cite{Wainwright:2019},
serves as a means to infer several statistical properties. 
Due to the uniform convergence property of the Glivenko-Cantelli estimate, at
the server there exists an implicit minimization of an empirical loss function
involving the actual and estimated CDFs, i.e. the objective function,
\begin{align}
	\|\widehat{F}_N(x) - F(x)\|_{\infty},
\end{align}
is minimized by when $\widehat{F}_N(x)$ is the Glivenko-Cantelli estimate of
$F(x)$. This minimization task is in agreement with the definition of federated
optimization~\cite{McMahan:2017}.
We discuss further applications of CDF estimation based signal analytics in the
section~\ref{sec:exp}.
\subsection{Federated learning algorithm for over-predictive signal analytics}
%
We consider the FL model where client devices work in a distributed manner. It
is assumed the each device has the hardware capability to compute the envelope
signal with the constraint, $\widehat{f}_i(t) \geq f_i(t)$ for $i \in
\{1,2,\ldots, D\}$. The following algorithm is proposed to obtain a
bandwidth-$L$ envelope approximation $\widehat{f}_i(t)$ and later aggregated to
derive a global signal analytic,
$G_{\text{server}}(\widehat{f}_1(t),\cdots,\widehat{f}_D(t))$.

%
\begin{algorithm}
  \caption{\label{alg:Fed_Learn} Federated learning algorithm for overpredictive signal analytics}
  \begin{algorithmic}
	  \STATE {\bfseries Input:} Recorded signals $f_i(t)$ at individual devices $i \in \{1,2,\ldots, D\}$.
	  \STATE {\bfseries Initialize:} Fourier coefficients -- $a_i[n]$ for $n \in \{-L, \ldots, 0, \ldots, L\}$ 
    \FOR{$i \in \{1,2,\ldots,D\}$}
    \STATE \begin{itemize}
	    \item[--] Compute $\widehat{f}_{i,\env}(t)$, and coefficients $b_i[n]$ for $n \in \{-L,\ldots, 0, \ldots, L\}$ such that $\widehat{f}_{i,\env}(t) \geq f_i(t)$.
	    \item[--] Communicate $b_i[n]$ to the server
    \end{itemize}	
    \ENDFOR
	  \STATE {\bfseries Return}  global model computed
	  $G_{\mbox{\footnotesize server}}\left(\widehat{f}_{1,\env},
	  \widehat{f}_{2,\env},\ldots,\widehat{f}_{D,\env}\right)$
	   (for example, \eqref{eq:AGG} or \eqref{eq:GCC}) 
  \end{algorithmic}
\end{algorithm}

In summary, the client step and the server step of the algorithm can be
explained as: 
\begin{enumerate}
	\item From the signal $f_i(t)$ recorded at device $i$, compute
		its envelope approximation $\widehat{f}_{i,\env}(t)$, using the
		envelope optimization \eqref{eq:sa1sa2}, and communicate the
		$(2L+1)$ Fourier coefficients to the server.
	\item At the server the Fourier coefficients received from each client
		device are aggregated, to calculate a global model (or
		statistic) of interest, denoted as $G_{\mbox{\footnotesize
		server}}\left(\widehat{f}_{1,\env},
		\widehat{f}_{2,\env},\ldots,\widehat{f}_{D,\env}\right)$.
%
%
\end{enumerate}
%

Step~1 of the algorithm executed at the client devices is outlined next. For
${\cal L}_1$ distance (see~\eqref{eq:L_1}), the optimization program becomes,
\begin{align}
\mbox{minimize } & b[0] - a[0] \nonumber \\
  \mbox{subject to } & \vec{b}^{\;T} \Phi(t) \geq f(t), 
\end{align}
where $\Phi(t) = [\exp(-2\pi L t), \ldots, \exp(2 \pi Lt)]^T$ and $\vec{b} =
(b[-L], \ldots, b[L])^T$ are the Fourier series coefficients of the envelope
approximation. It is known that the above linear program with linear
constraints is solvable efficiently~\cite{Kumar:2017}. For ${\cal L}_2$ case,
the cost function $b[0] - a[0]$ is replaced by the quadratic cost
in~\eqref{eq:L_2}.

We understand that, as approximation bandwidth $L$ is increased, the envelope
signal $\widehat{f}_{i, \env}(t)$ become more proximal to their target signal
$f_i(t)$. It is expected that the approximation errors, $\SA_1$ and $\SA_2$
will decrease as $L$ increases.  However, analyzing the dependence of $\SA_q, q
= 1, 2$ versus $L$ is difficult.  Accordingly, we rely on a na\"{i}ve envelope
approximation to analyze the fundamental bounds on their tradeoff.

\section{Performance Bounds on Optimal Envelope Approximation\label{sec:Main_results}}
In this section, we analyze the performance of the $L$ bandwidth envelope
approximation. In particular, we provide an upper bound for the envelope
approximation error, using the idea of na\"{i}ve envelope approximation scheme,
described below.
\subsection{A Na\"{i}ve Envelope Approximation Scheme}
\label{subsec:DC_boost}
Let us consider a single client device, $i=1$ in isolation.  Let
$f_{1,\proj}(t)$ be the orthogonal projection of $f_1(t)$ on the span of
$\exp{(j2\pi kt)}$ for $|k| \leq L$. Then $f_{1,\proj}(t) = \sum_{|k|\leq L}
a_1[k] \exp({j2\pi kt}).$ The na\"{i}ve envelope approximation scheme is as
follows~\cite{Maheshwari:2015}:
\begin{align} 
f_{1,\env}(t) = f_{1,\proj}(t) + C_0, \label{eq:naive_app}
\end{align}
where $C_0 = \|f_1 - f_{1, \proj}\|_\infty$. Using the triangle inequality, 
\begin{align}
C_0  \leq \sum_{|k|>L} |a_1[k]| \leq \sum_{|k|>L} \frac{C}{|k|^p}, \qquad p>1.
\label{eq:C0bound}
\end{align}
For $p > 1$~\cite[Sec.~2.2]{Bhatia:1993}, we can show that, $C_0 =
O\left(\frac{1}{L^{p-1}}\right)$.
%
%
\subsection{Envelope Approximation Analysis}
In this section we seek the answer to the question -- \textit{Is there a class
of signals for which the na\"{i}ve envelope approximation is a good scheme?}
The goodness here can be measured using the mean squared error. The result
discussed below shows that there exist a certain class of signals for which the
na\"{i}ve approximation is order optimal to the optimal envelope scheme, while
using ${\cal L}_1$ or ${\cal L}_2$ norm minimization.

In the result stated below we use the notation $\SA_q$ to denote the optimal
envelope approximation error (see \eqref{eq:L_1}), and the notation $\SA'_q$ to
be the approximation error corresponsing to the na\"{i}ve envelope approximation
in~\eqref{eq:naive_app}, where $q\in \{1,2\}$ according to the ${\cal L}_1$ or
${\cal L}_2$ cost metric.

\begin{theorem}
%
%
\label{th:main}
%
	The approximation errors of the optimal envelope signal and the
	na\"{i}ve envelope signal, \textit{viz.} $\SA_q$ and $\SA'_q$
	respectively for $q \in \{1,2\}$, are shown to be order optimal for a
	specific class of $p$-times differentiable signals as stated below.
	\begin{enumerate}
		\item[(i)] The approximation error of the optimal ${\cal L}_2$
			envelope signal, $\SA_2$, is order optimal to the
			approximation error of the na\"{i}ve envelope signal,
			$\SA'_2$ for the class of $p$-times differentiable
			signals where Fourier coefficients,
			$|a_1[k]|=\frac{1}{|k|^p}$ for $|k|>L$.  That is,
			\begin{align}
			1 \leq \frac{\SA'_2}{\SA_2} \leq
				\left(1+\frac{1}{L}\right)^{2p-1}.  \nonumber
			\end{align}
		\item[(ii)] The approximation error of the optimal ${\cal L}_1$
			envelope signal, $\SA_1$ is the same as the
			approximation error of the na\"{i}ve envelope signal,
			$\SA'_1$ for the class of $p$-times differentiable
			signals with Fourier coefficients, $a_1[k] \geq 0$
			(non-negative) and $a_1[k]=a_1[-k]$ (symmetric) for all $k
			\in \zZ$. That is, 
			\begin{align} 
				\SA_1 = \SA'_1 = 2 \sum_{k>L} a_1[k]. \nonumber 
			\end{align}
	\end{enumerate}
\end{theorem}
The detailed proof is shown in~\underline{\ref{app:order_optimal}}.

\noindent \textit{Remark}: The above result is shown for one (that is $D=1$)
client device. For $D$ clients with a sum signal analytic at the server (that
is $\widehat{s}(t) := \sum_{i=1}^D \widehat{f}_{i}(t)$), $\SA_1$ as well as
$\SA'_1$ scale linearly with $D$. In contrast, $\SA_2$ and $\SA'_2$ will scale
quadratically with $D$. Thus in both ${\cal L}_1$ and ${\cal L}_2$ norm based
error, the ratio between the approximation errors will remain the same, as in
the $D= 1$ case discussed in the proof.

\subsection{CDF of the Envelope Signal Approximation}
The CDF is a useful signal analytic to determine various statistical functions,
especially in a distributed learning setting such as federated learning.
Functions of the CDF estimates are beneficial in estimating customer usage
statistics such as mean, median or peak demand in the electricity smart
metering illustrated in Fig.~\ref{fig:Block_Dia}. By invoking
Therorem~\ref{th:main}, we derive performance bounds on the estimated CDF
obtained from the envelope approximation algorithms (see Sec.~\ref{sec:Algo}).
Let $F_X(x)$ denote the CDF of the actual signal $X(t)$ and let
$F_{X_{\env}}(x)$ be the CDF of the envelope signal approximation,
$\widehat{X}_{\env}(t)$, with $(2L+1)$ Fourier coefficients. In the result
below, we characterize an upper bound on the pointwise difference between the
actual and the estimated CDFs, as stated below.
\begin{theorem}
	The pointwise difference between the CDF of the actual signal, $F_X(x)$ and the
	CDF of the envelope signal approximation, $F_{\widehat{X}_{\env}}(x)$, with
	$(2L+1)$ Fourier coefficient is bounded by,
	\begin{align}
		F_X(x) - F_{\widehat{X}_{\env}}(x) \leq \frac{C}{L^{\frac{2p-1}{3}}},
	\end{align}
	where $C=\left(4^{\frac13} + 2\times 4^{-\frac23}\right)\frac{f_{X,\max}^{2/3}}{(2p-1)^{1/3}}$ and $f_{X,\max} := \begin{aligned}\max_x \left[\frac{d F_X(x)}{d x}\right]\end{aligned}$.
\end{theorem}
The proof of this result is available in \underline{\ref{app:Comm_Accuracy_Tradeoff}}

\subsection{Effect of Subsampling on CDF estimate}

In practice, signals are obtained as discrete samples over a finite interval.
Thus, it is essential to understand how envelope approximation can be performed
over signal samples and evaluate the tradeoff between the envelope
approximation error and the sampling rate. In this connection, we will consider
the aspect of signal sampling while limiting the discussion to the ${\cal L}_2$
distortion. However, a similar analysis also extends to the ${\cal L}_1$
distortion metric. The envelope signal approximation problem posed in
\eqref{eq:sa1sa2} is reformulated to reflect the finite signal sample
assumption as described below. If $\vec{b}_{\opt}$ represents the minimizer of
${\cal L}_2$ distance metric applied in \eqref{eq:sa1sa2}, then we define, 
\begin{align}
	\vec{b}_{\app,n} & :=\argmin_{\vec{b}}  \sum_{|k|\leq L} |b[k]-a[k]|^2 + \sum_{|k|>L} |a[k]|^2 \nonumber \\
	& \text{subject to} \quad \widehat{f}_{\env}(t_n) \geq f(t_n), \quad \forall \; t_n \in \left\{0,\frac1n, \frac2n,\ldots, 1\right\}, \label{eq:discrete_L_2}
\end{align}
where $n$ is a positive integer denoting the number of signal samples. Also, we
define the approximately optimal envelope signal, $f_{\app,n}(t):=
\vec{b}_{\app,n} \Phi(t)$. The above formulation in \eqref{eq:discrete_L_2} has
only finite inequality constraints as against the infinite inequality
constraints in \eqref{eq:sa1sa2}. Due to the relaxation of the envelope
constraints attributed to signal sampling, the ${\cal L}_2$ cost of the approximately
optimum envelope is lesser than the optimum envelope. This difference in the
${\cal L}_2$ cost of the envelope error, for the infinite envelope constraint
and the sampled envelope constraint relaxation as characterized
in~\cite{Kumar:2017} is written as,
\begin{align}
	\sum_{|k| \leq L} \left[\left|b_{\opt}[k] - a[k]\right|^2 - \left|b_{\app,n}[k] - a[k]\right|^2\right] \leq 2 (b_{\app,n}[0] - a[0]) \left(\frac{c+c'}{n}\right) + o(1/n). \label{eq:discrete_bound}
\end{align}
The constants $c$ and $c'$ in the above upperbound expression corresponds
to the maximum slopes of the actual signal and the approximately optimal
envelope signal respectively. That is, $|f'(t)|\leq c$ and
$|f'_{\app,n}(t)|:=c'$. The upperbound on $c'$ is shown to be $2\pi L
\|f\|_{\infty}$, in \cite{Kumar:2017}. 

Let $F_X(x)$ be the CDF of the actual signal, $X(t):= \sum_{k \in \zZ}
A[k]\exp(j2 \pi k t)$ for $t \in [0,1]$.  Further, let $F_{X_{\app,n}}(x)$ be
CDF of the envelope signal approximation with the subsampled constraints,
$\widehat{X}_{\app,n}(t): = \sum_{|k| \leq L} B_{\app,n}[k] \exp(j 2 \pi k
t)$ for $t \in [0,1]$, where the set of coefficients $\{B_{\app,n,i}[k]: -L
\leq k \leq L\}$ is as defined in~\eqref{eq:discrete_L_2}. An upperbound
between the pointwise difference of the two CDFs is shown below, 
\begin{theorem}
	The pointwise difference between the CDF of the actual signal, $F_X(x)$
	and the CDF of the subsampled envelope signal approximation, $F_{X_{\app,n}}(x)$,
	with $(2L+1)$ Fourier coefficient is bounded by,
	\begin{align}
		F_X(x) - F_{X_{\app,n}}(x) \leq C \times \left\{\frac{4}{2p-1} \frac{1}{L^{2p-1}} + 8\mu_{\app,n}\frac{c+c'}{n} + o(1/n)\right\}^{1/3},
	\end{align}
	where $C = \left(2^{\frac13} + 2^{-\frac23}\right) f_{X,\max}^{2/3}$
	and $f_{X,\max} := \begin{aligned}\max_x \left[\frac{d F_X(x)}{d
	x}\right]\end{aligned}$. The constant, $\mu_{\app,n} =
	\eE\left[B_{\app,n}[0]-A[0]\right]$, where $B_{\app,n}[0]$ and $A[0]$
	are the zero frequency components corresponding to the signals
	$\widehat{X}_{\app,n}(t)$ and $X(t)$ respectively.
\end{theorem}
The proof of this result is available in~\underline{\ref{app:Subsampling_CDF}}.

\section{Experimental Results and Discussion}\label{sec:exp}
\subsection{Electricity Consumption Dataset}
\begin{figure}[t]
  \centering
  \includegraphics[scale=0.45]{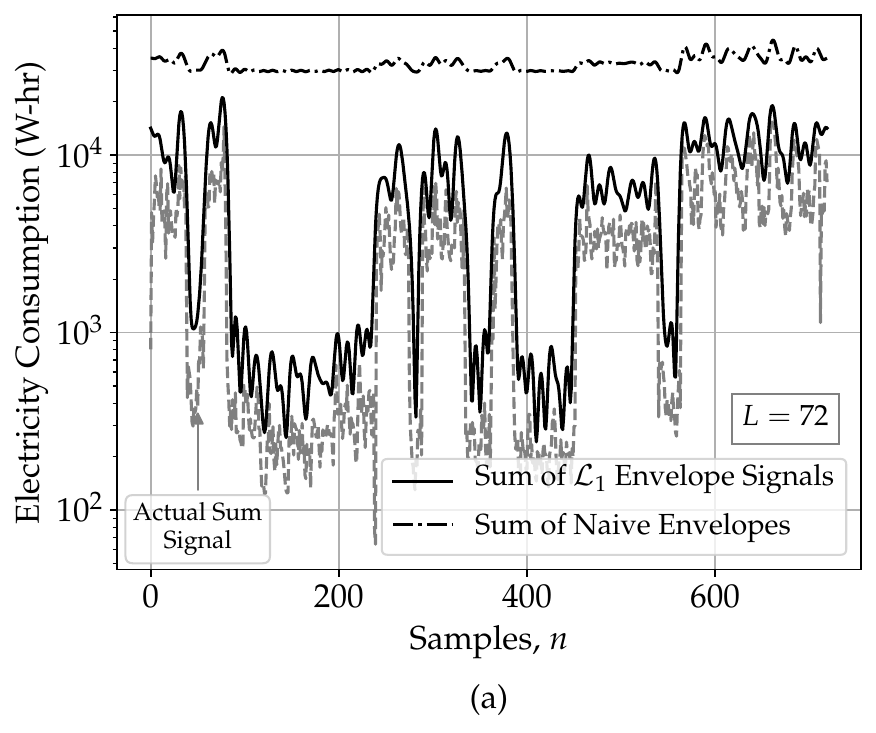}
  \includegraphics[scale=0.45]{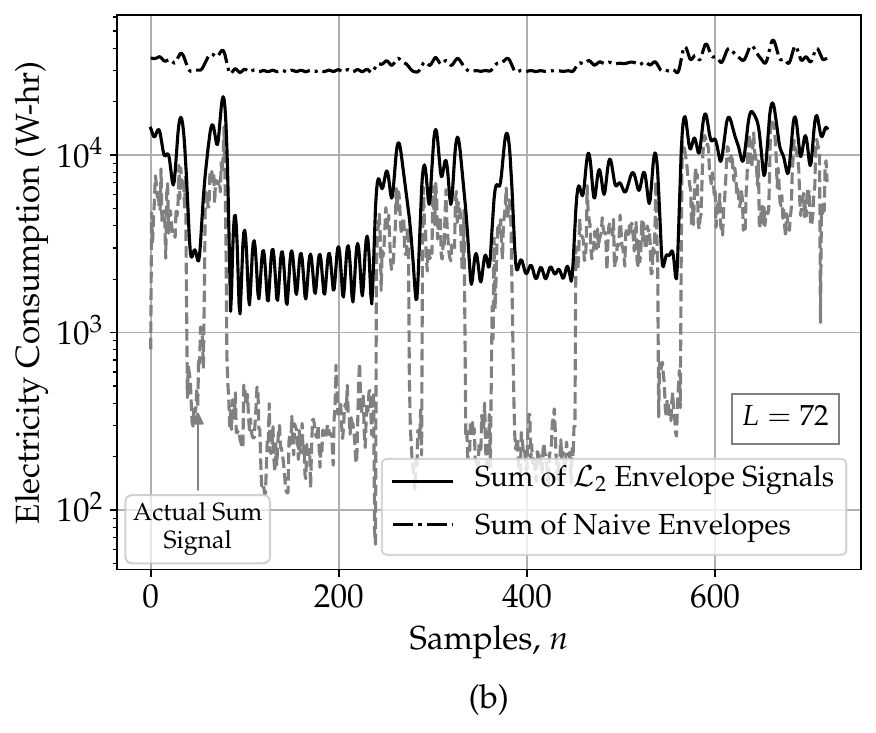}
  \caption{\label{fig:L2_Signal} The sum of client signals obtained by
	distributed envelope approximation schemes are compared with the ground
	truth signal, when the number of coefficients communicated are $L =72$.
	Plot (a) and (b) consider ${\cal L}_1$ and ${\cal L}_2$ cost functions
	respectively. It is observed that the optimal envelope approximation
	signal estimates the peak energy demand regions.}
\end{figure}
\begin{figure}[t]
  \centering
  \includegraphics[scale=0.55]{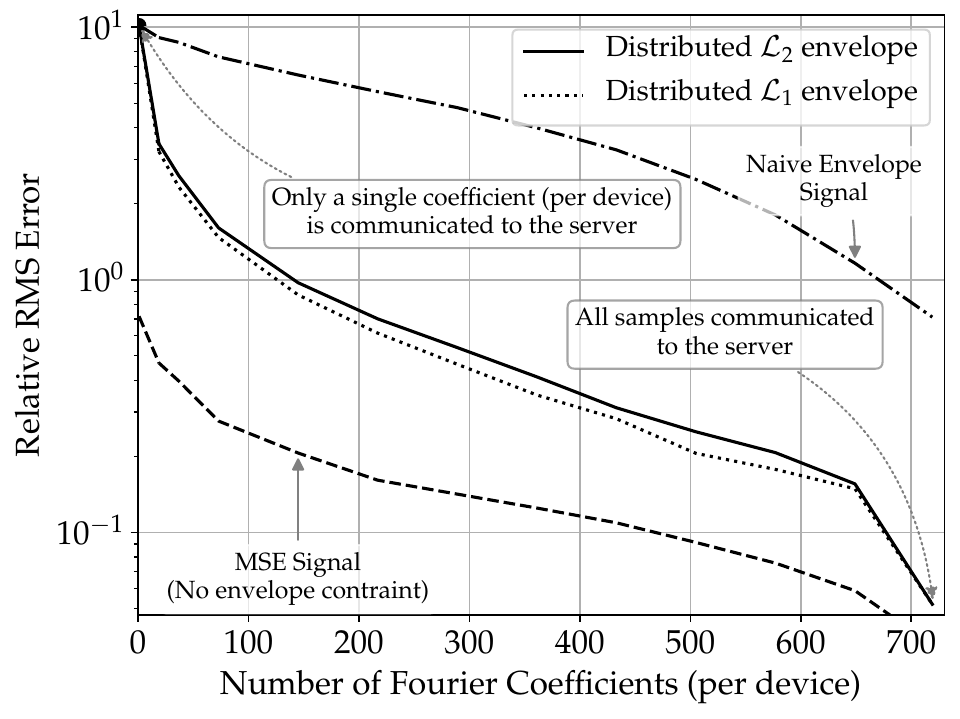}
  \caption{\label{fig:L2_Tradeoff} The depiction of the tradeoff between
	normalized root mean-squared (RMS) error and the number of Fourier coefficients
  communicated per device. The distributed envelope signal approximation based
	on ${\cal L}_1$ and ${\cal L}_2$-norms converge to the zero error with increase in
	the number of communicated signal coefficients. The plot also shows the RMS error for the  na\"{i}ve envelope signal and the MSE optimal signal (which is devoid of any envelope constraints).}
\end{figure}
%
%
To showcase the effect of envelope approximation in a federated learning model,
we have performed experiments on an existing electrical energy consumption
dataset~\cite{Mammen:2018}, consisting of hourly energy consumption of 39
users from a residential building. 
Users with atleast 30 days of synchronized timestamps are filtered for the
experiments discussed here -- which results in 37 users. In the considered
dataset, energy measurements are available in three phases. However, we have
restricted to only a single phase, namely W3 in the dataset. 

\vspace{1ex}
\textbf{Hardware/Software Configuration}: The simulations on the dataset were
performed in a PC with the Processor model -- Intel(R) Core(TM) i3-2310M
CPU @ 2.10GHz, 2100 Mhz, 2 Core(s), RAM 6 GB; and implemented in MATLAB 2015b
(Windows platform) using the standard curve fitting toolbox and CVX package (ver
2.1).

\subsection{Tradeoff between Communication Cost and Approximation Error}
We analyze the tradeoff between the number of communicated Fourier coefficients
and the envelope approximation error by considering a sum signal analytic (or
the sum of user energy consumptions, refer~\ref{eq:AGG}), which is learned at
the central server.
Each Fourier coefficient communicated to the server is a real-valued floating
point number typically 4 Bytes (or 32 bits).  
Thus, each client device (i.e., the 37 users) computes the Fourier coefficients
of the envelope approximation, which represents 30 days of hourly data, or 720
signal samples~(refer Fig.~\ref{fig:L2_Signal}(a)-(b) for the time-series plot
of the sum signal analytic). 
Fig.~\ref{fig:L2_Tradeoff} shows the tradeoff plot between the relative root
mean-squared (RMS) error and the number of Fourier coefficients. We observe a
graceful degradation of the RMS error with respect to the number of
coefficients, for the optimal envelope approximation, derived using the ${\cal
L}_1$ and ${\cal L}_2$ cost functions.
We note that the RMS error in the distributed signal approximation schemes
approach zero when 720 Fourier coefficients are transmitted. 
However, for the considered dataset, the naive approximation fails to capture
the envelope trend.  
In Fig.~\ref{fig:L2_Tradeoff}, we have indicated two extremum points -- one
corresponding to the least communication scheme(that is $L=1$), and the other
where all coefficients are communicated (that is $L=360$). 
The ${\cal L}_1$ cost function results in a relatively lower RMS error compare
to ${\cal L}_2$, which is attributed to the better time-series fit of ${\cal
L}_1$ as shown in Fig.~\ref{fig:L2_Signal}. 
Further, it can be observed that the relative approximation error of classical
MSE minimizer (that is, without the envelope constraint) is lower than optimal
${\cal L}_1$ and ${\cal L}_2$ distributed envelopes by one order of magnitude.
\subsection{Signal Analytics based on Cumulative Distribution Function}

\begin{figure}[t]
  \centering
  \includegraphics[scale=0.45]{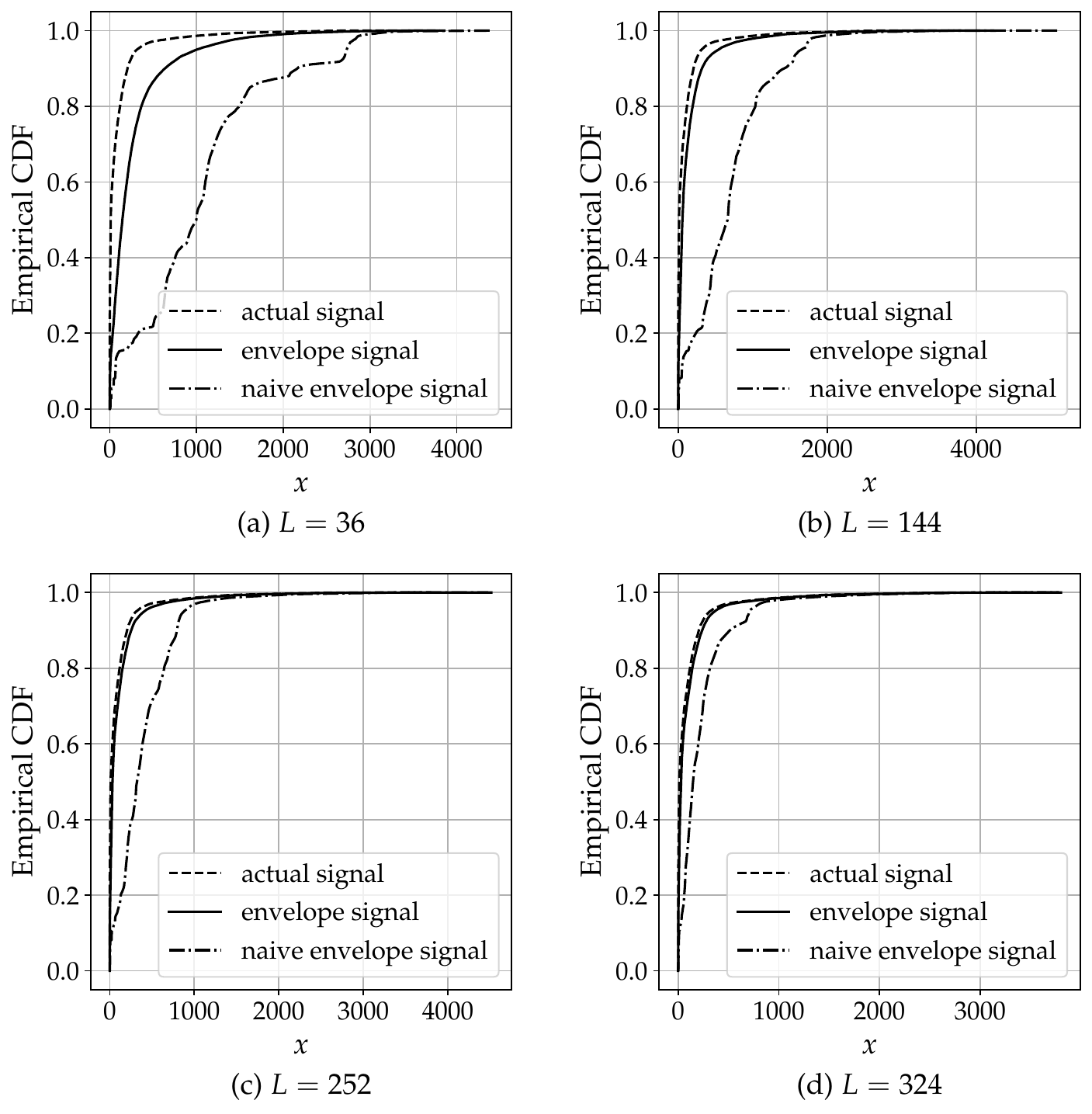}
  \caption{\label{fig:L2_CDFs} Plot shows the empirical CDF approximation at
	different approximation levels $L$ and illustrates its convergence to
	the actual CDF for the ${\cal L}_2$ cost function. Note that the
	optimum and the na\"{i}ve envelope signal CDFs always appear on the
	right of the actual CDF due to the envelope constraint.}
\end{figure}

In this experiment, we study the signal analytics learned from the CDF of the
reconstructed envelope signals. At the central server, we compute empirical
CDFs using the Glivenko Cantelli estimate, which are later processed to infer
statistical properties involving symmetric function.  Fig.~\ref{fig:L2_CDFs}
illustrates the convergence of the estimated CDF (from envelope approximations
using the ${\cal L}_2$ cost) to the baseline CDF, which is constituted by all
the raw data observed at the clients. For lower approximations, e.g.  $L=36$,
the CDF estimate performs poorly for the mid-signal ranges ($x \in [200,1500]$)
while being able to accurately track the signal peaks ($x \in [2000, 4000]$).
The gap between the estimate and the baseline reduces with an increase in
approximation coefficients, $L$. We note that the overpredicted CDF always
appears towards the right of the true CDF due to the envelope constraint.

Next, we attempt to infer quantile estimates from these CDF plots. In
Table~\ref{tab:quantile}, we compare the quantiles of the actual CDF with the
quantiles of the CDF estimates. 
We note that $\mathcal{L}_1$ based quantile estimates provide a closer
approximation to the actual signal quantile as when compared to $\mathcal{L}_2$
scheme. Also, the higher-order quantiles deviate much compared to the lower
ones, which is attributed to the overprediction constraint at the devices. For
comparison, we have also included the quantiles that correspond to the
na\"{i}ve envelope signal. By choosing an appropriate basis representation, we
can improve the accuracy of the quantile estimates in Table~\ref{tab:quantile}.
We shall address this in a future work.
\begin{table}[!hbt]\scriptsize
  \centering
  \caption{\label{tab:quantile} A comparison of the quantiles of the true signal
  with the envelope approximation signal}
  \renewcommand{\arraystretch}{1.5}
  \begin{tabular}{@{}cccccclccc@{}}
    \toprule
    \multicolumn{1}{c}{\multirow{2}{*}{\bf Quantile}} &
	  \multicolumn{1}{c}{\multirow{2}{*}{\parbox{1cm}{\centering \bf  Actual Signal}}} &
	  \multicolumn{1}{c}{\multirow{2}{*}{\parbox{1.5cm}{\centering \bf Cost Function}}} &
	  \multicolumn{3}{c}{\centering \bf Envelope Signal} & &  \multicolumn{3}{c}{\centering \bf Na\"{i}ve Envelope Signal} \\
	  \cline{4-6} \cline{8-10}
	  \multicolumn{1}{c}{} & \multicolumn{1}{c}{} & \multicolumn{1}{c}{} & $L = 36$ & $L = 180$ & $L =324$ & & $L = 36$ & $L = 180$ & $L = 324$\\
    \midrule 
	  \multirow{2}*{$\mathbf{10\%}$} & \multirow{2}*{$0$} & ${\cal L}_1$ & $0.21$ & $10^{-6}$ & $10^{-7}$ & & \multirow{2}{*}{$66.34$} & \multirow{2}{*}{$39.19$} & \multirow{2}{*}{$7.79$}\\ \cline{3-6} & & ${\cal L}_2$ & $3.39$ &  $0.80$ & $0.11$ & & & &  \\
	  \midrule
	  \multirow{2}*{$\mathbf{50\%}$} & \multirow{2}*{6.09} &  ${\cal L}_1$ & $19.25$ & $10.12$ & $6.78$ & & \multirow{2}{*}{$754.85$} & \multirow{2}{*}{$410.91$} & \multirow{2}{*}{$116.05$} \\ \cline{3-6} & &  ${\cal L}_2$ & $65.36$ & $27.55$ & $13.36$  & & & & \\
	  \midrule
	  \multirow{2}*{$\mathbf{90\%}$} & \multirow{2}*{204.53} & ${\cal L}_1$ & $429.59$ & $272.41$ & $225.71$ & & \multirow{2}{*}{$2162.7$} & \multirow{2}{*}{$1229.2$} & \multirow{2}{*}{$514.11$}\\ \cline{3-6}  & & ${\cal L}_2$ & $431.21$ & $268.6$ & $218.42$ & & & & \\
    \bottomrule
  \end{tabular}
\end{table}

\subsection{Effect of Subsampling}
\begin{figure}[!hbt]
  \centering
  \includegraphics[scale=0.5]{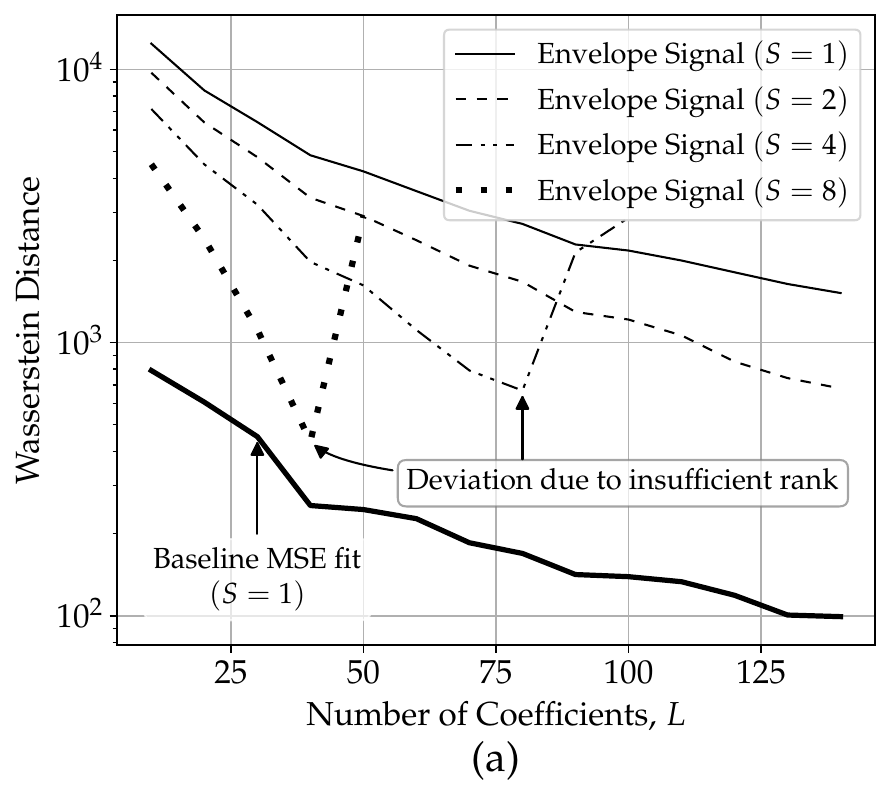}
  \includegraphics[scale=0.5]{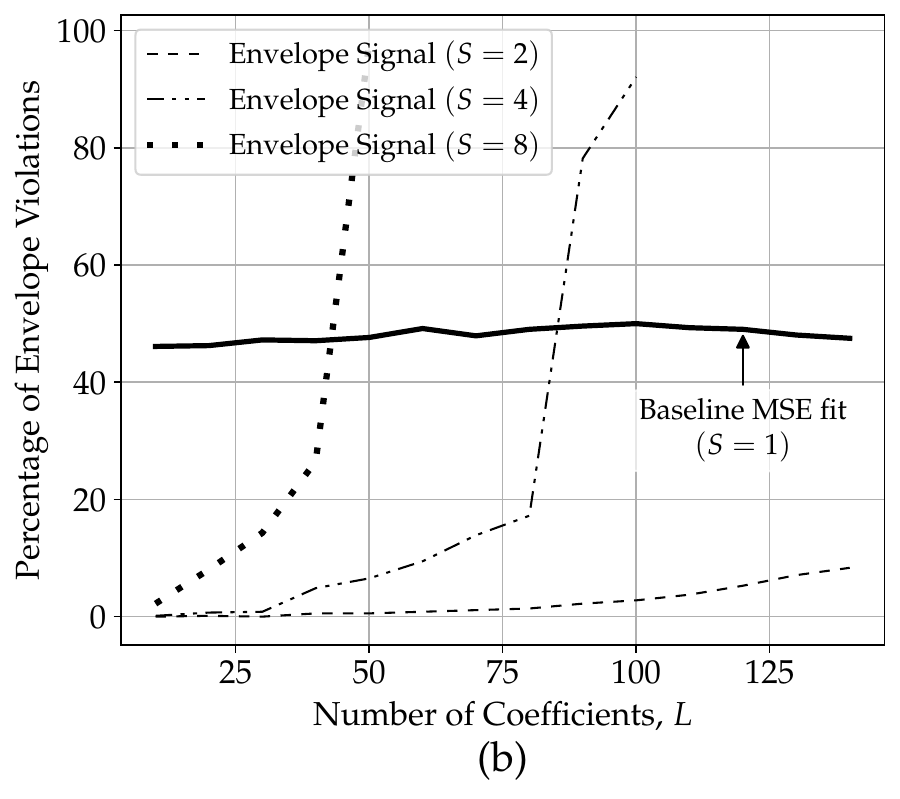}	
  \includegraphics[scale=0.5]{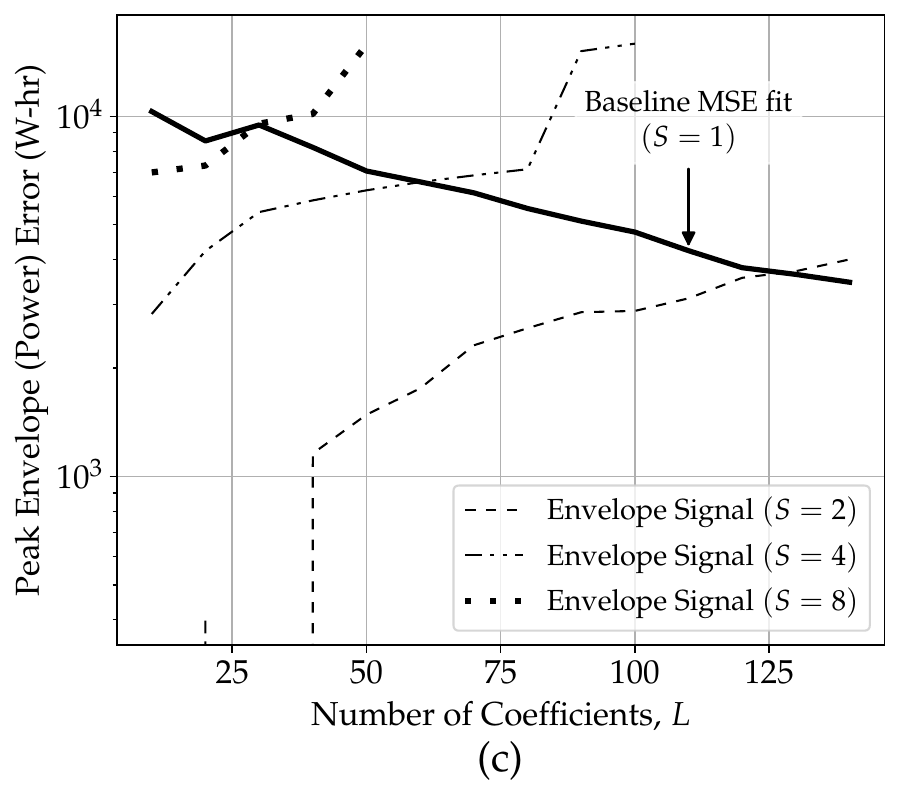}	
	\caption{\label{fig:Subsamp-figs} The figures depict the effect of subsampling on the ${\cal L}_1$ cost based envelope approximation based on three performance metrics: (a) Wasserstein distance of estimated envelope CDF and true CDF, (b) percentage of envelope violations at different subsampling rates, (c) peak envelope error due to subsampling.}
\end{figure}

In practice, as the signal acquisition in the edge device will not be a
recording a continuous signal, we will investigate the effect of signal
sampling on the envelope approximation algorithms. Since the dataset considered
here is already discrete time-series with $720$ samples, we will see the effect
of subsampling on the various performance metrics. In particular, for this
experimental study we consider three performance error metric, \textit{viz.}
(i) Wasserstein distance between the estimated envelope CDF and the actual CDF,
(ii) Number of envelope constraint violations with respect to the actual $720$
samples, and (iii) Peak envelope error after subsampling. We have used the 1-D
Wasserstein distance between the CDF of the envelope signal and the CDF of the
actual signal, that is defined as,
\begin{align*}
	{\cal W}(X,\widehat{X}_{\env}) := \int_0^1\left|F_{X}^{-1}(z) - F_{\widehat{X}_{\env}}^{-1}(z)\right| dz,
\end{align*}
where $F_{\widehat{X}_{\env}}(x)$ is the envelope signal CDF and $F_{X}(x)$ is
the true signal CDF. The number of envelope constraint violations and the peak
envelope error are computed based on the pointwise difference between the sum
of envelope approximated time-series and the sum of actual time-series. In
Fig~.\ref{fig:Subsamp-figs}, we show the various performance metrics
considering the ${\cal L}_1$ cost function for different subsampling rate,i.e. $S
\in \{1,2,4,8\}$. At a subsampling rate of $S$, the number of envelope
constraints is $720/S$. From Fig.~\ref{fig:Subsamp-figs}~(a), it is observed
that subsampling by considering alternate signal samples, i.e. $S=2$, reduces
the Wasserstein distance between the CDFs of the envelope signal and the true
signal, when compared with the envelope without subsampling, i.e. $S=1$. This
is attributed to the envelope constraint relaxation upon subsampling, which
results in a lesser minima for the same objective function. Since the CDF of
the MSE approximation signal does account for the envelope constraints the
Wasserstein distance approach zero with increasing approximation coefficients,
as shown in Fig.~\ref{fig:Subsamp-figs} (a). 

However, subsampling introduces violations of the envelope constraint as shown
in Fig.~\ref{fig:Subsamp-figs}~(b)-(c). Both the percentage of envelope
constraint violations and the peak error due to envelope constraint violation
increase with the number of approximation coefficients $L$. This is because of
the effect of signal aliasing and signal overfitting introduced due to
subsampling. Further it is noted that the signal reconstruction from subsampled
time-series fails when the number of Fourier coefficients is more than the
total number of signal samples, i.e. $2L+1 \geq 720/S$. In this case envelope
approximation algorithm fails to solve the optimization program owing to rank
deficiency. This effect is shown in  Fig.~\ref{fig:Subsamp-figs} (a), where the
Wasserstein distance returned by the solver starts to increase at $L=45$ for
$S=8$ and $L=90$ for $S=4$. For $S=2$, the rank violation will begin at $L=180$
although not shown in the plot.

The rank deficiency also results in an abrupt increase in envelope errors as
depicted in Fig.~\ref{fig:Subsamp-figs} (b)-(c). The tradeoff
between the sampling rate, Wasserstein's distance and the envelope violations,
suggests that envelope approximation algorithm should be limited to $2L+1 <
720/S$, where the aliasing effects due to subsampling are small. A similar
trend is seen to hold when the ${\cal L}_1$ cost function is replaced by the
${\cal L}_2$ cost function. We make a comparison between the envelope
approximation algorithms of ${\cal L}_1$ and ${\cal L}_2$ cost functions based
on the three performance metrics for $S=2$ in Table~\ref{tab:envViola}. Due to
the relatively poor signal fitting of the ${\cal L}_2$ compared to ${\cal L}_1$
as shown in Fig.~\ref{fig:L2_Signal}, the Wasserstein distance is more for
${\cal L}_2$ cost. However the number of envelope errors and the peak envelope
error is reduced because of the higher envelope constraint guard region at
lower signal amplitudes for the ${\cal L}_2$ approximation. Thus, for the low
envelope approximation regime, i.e.  $L=10$ to $L=50$, it is suggestive to
employ the ${\cal L}_2$ approximation algorithm when subsampling, as the
envelope constraint violations are better controlled. 

%
%
%
\begin{table}[!hbt]\scriptsize
  \centering
	\caption{\label{tab:envViola} Comparison of different error metric for the ${\cal L}_1$ and the ${\cal L}_2$ cost functions, assuming $S=2$ subsampling}
  \renewcommand{\arraystretch}{1.5}
	\begin{tabular}{@{}cccccccccc@{}}
    \toprule
	\multicolumn{1}{c}{\multirow{2}{*}{\parbox{2cm}{\centering\bf
		Error Metric}}} & \multicolumn{1}{c}{\multirow{2}{*}{\parbox{1.5cm}{\centering\bf
		Cost Function}}} & \multicolumn{8}{c}{ \centering \bf Envelope
		Approximation}  \\ 
	\cline{3-10} 
		\multicolumn{1}{c}{} & \multicolumn{1}{c}{} & $L=10$ & $L=20$ & $L=30$ & $L=40$ & $L=50$ & $L= 60$ & $L=70$ & $L=80$ \\ 
	\midrule 
		\multicolumn{1}{c}{\multirow{2}{*}{\parbox{2.5cm}{\centering\bf Wasserstein Distance}}} & ${\cal L}_1$ & $9725.63$ & $6396.24$ & $4775.37$ & $3390.32$ & $2906.97$ & $2370.42$ & $1913.22$ & $1670.10$ \\
		& ${\cal L}_2$ & $11378.52$ & $8051.19$ & $6050.24$ & $4403.93$ & $3867.93$ & $3220.06$ & $2725.07$ & $2400.25$ \\	
	\midrule
		\multicolumn{1}{c}{\multirow{2}{*}{\parbox{2.5cm}{\centering\bf Number of Envelope Violations}}} &  ${\cal L}_1$ &  $0$ & $1$ & $0$ & $4$ & $4$ & $6$ & $8$ & $10$ \\
		&  ${\cal L}_2$ & $0$ & $0$ & $0$ & $2$ & $2$ & $5$ & $6$ & $7$ \\	
	\midrule
		\multicolumn{1}{c}{\multirow{2}{*}{\parbox{2.5cm}{\centering\bf Peak Envelope Error (in W-Hr)}}} &  ${\cal L}_1$ & $0$ & $408.26$ & $0$ & $1158.47$ & $1484.85$ & $1756.70$ & $2313.55$ & $2581.35$ \\
		& ${\cal L}_2$ & $0$ & $0$ & $0$ & $683.21$ & $793.70$ & $1271.43$ & $2044.13$ & $2363.26$ \\
		%
	\bottomrule
	\end{tabular}

\end{table}

\section{Conclusions}
In this paper, we proposed signal approximation algorithms to perform
overprediction of distributed signals in a client-server federated learning
model. Using a convex optimization framework, we determine overpredicting
signal representations locally at each edge device, further communicating these
to the cloud server to compute signal analytics. Such signal analytics, like
the aggregate signal or the Cumulative Distribution Function (CDF) of the
time-series signal, derived from the individual signal representations using
federated aggregation, were analyzed to determine the tradeoff characteristics
between the mean-squared approximation error and the number of approximation
coefficients (communicated bytes). Particularly for the CDF approximation, we
used the Glivenko-Cantelli estimate to estimate upperbounds on the pointwise
difference between the actual CDF and the approximate CDF. Experiments on a
residential energy consumption dataset validate the analytical performance
bounds of the envelope signal analytics considered. In the future, we envisage
extending the proposed algorithms for time-series prediction considering the
signal overprediction constraints.
%

\appendix
\section{Order Optimality of Envelope Approximation}\label{app:order_optimal}
\begin{proof} :  
	We show the desired result for $q=2$ and $q=1$ separately. For
	notational simplicity we will consider the following analysis with
	respect to client device $1$ with observed signal being $f_1(t)$ and
	the corresponding Fourier coefficients as $\{a_1[k] : k\in \zZ\}$. The
	envelope approximation and its Fourier coefficients are represented as
	$\widehat{f}_{1,\env}(t)$ and $\{b_1[k]: -L\leq k\leq L\}$ respectively.

	\underline{\textbf{Case 1}}: For $q=2$, the optimal envelope
	approximation error is computed from the envelope constraint
	$\vec{b}_1^T\Phi(t) \geq f_1(t)$. Since $f_1(t) = \sum_{k\in \zZ} a_1[k]\exp{(j
	2\pi kt)}$ for $t \in [0,1]$, we have the bound
	\begin{align}
		\sum_{|k|\leq L} b_1[k] \exp{(j 2\pi kt)} & \geq \sum_{k\in \zZ} a_1[k]\exp{j 2\pi kt} \nonumber \\
		\Rightarrow \sum_{|k|\leq L} (b_1[k]-a_1[k])\exp{(j 2\pi kt)} & \geq \sum_{|k| > L} a_1[k]\exp{j 2\pi kt}, \nonumber 
	\end{align}
	that leads to the inequality $\sum_{|k|\leq L} |b_1[k]-a_1[k]|^2 \geq
	\sum_{|k|>L} |a_1[k]|^2$, based on the Parseval's Theorem. Now, the
	approximation error of the optimal ${\cal L}_2$-norm envelope
	approximation signal,
	\begin{align}
		\SA_2&:=\min_{\vec{b}_1} \sum_{|k| \leq L} |b_1[k] -a_1[k]|^2 + \sum_{|k|>L}
	|a_1[k]|^2 \nonumber \\
		& \geq \min_{\vec{b}_1} \quad 2 \sum_{|k|>L} |a_1[k]|^2.
	\end{align}
	Further, using Fact~\ref{lem:regularity} for $p$-times differentiable
	signal class with $a_1[k]\;, k \in \zZ$ satisfying the lower
	bound~\cite{Bhatia:1993} for the series $\sum_{|k|>L} \frac{1}{|k|^{2p}}$, 
	\begin{align}
		\SA_2 \geq \frac{2}{2p-1}\frac{1}{(L+1)^{2p-1}}. \label{eq:SA_2}
	\end{align}
	Using \eqref{eq:C0bound}, the approximation error of the na\"{i}ve
	envelope can be upper bounded as~\cite{Bhatia:1993},
	\begin{align}
		\SA'_2 \leq \frac{2}{2p-1}\frac{1}{L^{2p-1}}. \label{eq:SA_2'}
	\end{align}
	Since $\SA_2$ refers to the approximation error of the optimal
	envelope, $\SA_2 \leq \SA_2'$, and we get the inequality,
	\begin{align}
		1 \leq \frac{\SA_2'}{\SA_2} \leq
		\slfrac{\left[\frac{2}{2p-1}\frac{1}{L^{2p-1}}\right]}{\left[\frac{2}{2p-1}
		\frac{1}{(L+1)^{2p-1}}\right]}, 
	\end{align}
	or $1 \leq \frac{\SA_2'}{\SA_2} \leq
	\left(1+\frac{1}{L}\right)^{2p-1}$. The right hand upperbound
	approaches $1$ in the limit $L \rightarrow \infty$. This shows that the
	approximation errors of the optimal envelope and the na\"{i}ve envelope
	are order optimal for the ${\cal L}_2$ cost.

%
	\underline{\textbf{Case 2}}: The optimality of the na\"{i}ve envelopes
	for the $\SA_1$ distance holds for a signal class with the following
	additional properties: 
	\begin{enumerate}
		\item[(i)] the Fourier coefficients $a_1[k] \geq 0$,
		\item[(ii)] $f_1(t)$ is real and even, that is $a_1[k] = a_1[-k]$.
	\end{enumerate}
	From these symmetry assumptions, it follows that $b_1[k] = b_1[-k]$.
	Restricted to this signal class, the $\SA_1$ envelope approximation is
	re-stated as:
\begin{align}
	\SA_1:= \min_{b_1[k], |k| \leq L} b_1[0] - a_1[0] &, \quad \text{subject to}  \nonumber \\
b_1[0] + 2 \sum_{ 1 \leq k \leq L}  b_1[k] \cos(2\pi k t)  & \geq \sum_{k \in \zZ}
a_1[k] e^{j 2\pi k t} \label{eq:envapp_L1}
\end{align}
  The above optimization can be shown to result in $b_{1,\text{\scriptsize opt}}
  [0] -a_1[0]= \sum_{|k|>L} a_1[k]$, using the following argument. Consider the
  envelope constraint in \eqref{eq:envapp_L1}, by rearranging the terms,
  \begin{align}
	  b_1[0] -a_1[0] \geq 2 \sum_{1\leq k \leq L} (a_1[k]-b_1[k]) \cos(2\pi k t) + 2 \sum_{|k| > L}
	  a_1[k] \cos(2\pi k t). 
  \end{align}
  Since we assumed $a_1[k]>0$ for all $k$ and $\{b_[k]: 1\leq k \leq L\}$ is
  the optimization variable, the right hand side term of the above inequality
  is maximum at $t=0$ and $t=1$. Thus, by choosing $b_1[k]= a_1[k]$ for $1\leq k
  \leq L$, we get $\SA_1:=b_{1,\opt}[0]-a_1[0] \geq 2\sum_{k >L} a[k]$. For the
  considered signal class, the na\"{i}ve approximation error, $\SA_1'$ is shown to satisfy 
  $\SA_1'= C_0 = \|f_1 - f_{1,\proj}\|_\infty \leq 2\sum_{k>L} a_1[k]$ using
  \eqref{eq:C0bound}. Since the optimal envelope has the minimum approximation
  error, $\SA_1 \leq \SA'_1$. In summary, 
  \begin{align}
	  2\sum_{k >L} a_1[k] \leq \SA_1 \leq \SA'_1 \leq 2\sum_{k >L} a_1[k],
  \end{align}
  which shows that $\SA_1 = \SA'_1$ for the signal class considered.

	\textit{Remark}: From \eqref{eq:L_inf}, the $L_\infty$ norm envelope
	distortion is expressed as an upper bound in terms of the Fourier
	coefficients.  For the $p$-times differentiable signal class, this
	results in a distortion, $\SA'_\infty= {\cal
	O}\left(\frac{1}{L^{p-1}}\right)$, on the na\"{i}ve approximation
	scheme, as expressed in Table~\ref{tab:UB_LB}.  We observe that an
	analytical expression for $\SA_\infty$ is challenging to determine
	without additional information on the signal class.  
\begin{table}[!htb]
\centering
\caption{\label{tab:UB_LB} Bounds on the approximation errors}
  \bgroup \def\arraystretch{1.35} \begin{tabular}{@{}lcc@{}}
%
\toprule
$\mathcal{L}_q$ norm & $\SA_q$  & $\SA'_q$    \\[1.2ex] \midrule
  $q = 1$  & $b_1[0] - a_1[0]$ &  $\begin{aligned}\sum_{|k|>L}
  |a_1[k]|\end{aligned}$ 
\\[3ex] 
	  $q = 2$ & $\begin{aligned} \frac{2}{2p-1}\frac{1}{(L+1)^{2p-1}}\end{aligned}$
		  &  $\begin{aligned} \frac{2}{2p-1}\frac{1}{L^{2p-1}} \end{aligned}$ \\[3ex] 
	  $q = \infty$ & $-$ & $\begin{aligned} \frac{2}{p-1}\frac{1}{L^{p-1}} \end{aligned}$ \\[3ex]
	  \bottomrule
\end{tabular} \egroup
\end{table}
\end{proof}
\section{Communication vs Accuracy Tradeoff in Envelope CDF Estimation~\label{app:Comm_Accuracy_Tradeoff}}
\begin{proof}
	Using the result from Grimmett and Stirzaker~\cite{Grimmett:2001}, for any $\epsilon >0$;
	\begin{align}
		F_{X_{\env}}(x) & \geq F_X(x-\varepsilon) - \pP\left(X_{\env}-X > \varepsilon \right) \nonumber \\
		\Rightarrow F_X(x) - F_{X_{\env}}(x) & \leq F_X(x) - F_X(x-\varepsilon) + \pP\left(X_{\env}-X > \varepsilon \right) \label{eq:Diff-CDF}
	\end{align}
	In the limit $\varepsilon \rightarrow 0$, 
	\begin{align}
		F_{X}(x) - F_{X_{\env}}(x) \leq \varepsilon f_X(x) +
		\pP\left(X_{\env}-X > \varepsilon \right) 
	\end{align}
	Since $X_{\env}$ is the $(2L+1)$ coefficient envelope approximation,
	\begin{align}
		\|X - X_{\env}\|_2^2 = \sum_{|k|\leq L} \left|B_{\env}[k]-A[k]\right|^2 + \sum_{|k|>L} |A[k]|^2 \label{eq:L2-sig-diff}
	\end{align}
	From the envelope constraint, $X_{\env}(t) \geq X(t)$, we have 
	\begin{align}
		\sum_{|k|\leq L} B_{\env}[k] e^{j2\pi kt} & \geq \sum_{k \in \zZ} A[k] e^{j2\pi kt}\nonumber \\
		\Rightarrow \sum_{|k|\leq L} \left(B_{\env}[k]-A[k]\right) e^{j2\pi kt} & \geq \sum_{|k| > L} A[k] e^{j2\pi kt} \nonumber \\
		\Rightarrow \sum_{|k|\leq L} \left|B_{\env}[k]-A[k]\right|^2  & \geq \sum_{|k| > L} |A[k]|^2
	\end{align}
	Applying the above inequality in \eqref{eq:L2-sig-diff},
	\begin{align}
		\|X_{\env} - X\|^2 & \leq 2 \sum_{|k|\leq L}\left|B_{\env}[k]-A[k]\right|^2 \nonumber \\
		& \leq 2 \sum_{|k|\leq L}\left|B_{\text{\footnotesize na\"{i}ve}}[k]-A[k]\right|^2 \nonumber \\
		& \leq \frac{2}{2p-1} \frac{1}{L^{2p-1}} \quad \text{ (Ref:~\cite{Bhatia:1993})}.
	\end{align}
	We now use the Chebyshev's inequality in \eqref{eq:Diff-CDF}, i.e.
	\begin{align}
		\pP(X_{\env}-X>\varepsilon) & \leq \frac{1}{\varepsilon^2} \eE\left(\|X_{\env}-X\|^2\right) \nonumber \\
		& \leq \frac{1}{\varepsilon^2} \frac{2}{2p-1}\frac{1}{L^{2p-1}}
	\end{align}
	The difference of the CDF's,
	\begin{align}
		F_X(x) - F_{X_{\env}}(x) \leq \varepsilon f_X(x) + \frac{2}{\varepsilon^2(2p-1)}\frac{1}{L^{2p-1}}.
	\end{align}
	Assuming $f_X(x) \leq f_{X,\max}$, we get $F_X(x) - F_{X_{\env}}(x) \leq \varepsilon f_X(x) + \frac{2}{\varepsilon^2(2p-1)}\frac{1}{L^{2p-1}}$. On minimizing the right-hand side w.r.t $\varepsilon$, we get,
	\begin{align}
		F_X(x) - F_{X_{\env}}(x) \leq \left(4^{\frac13} + 2\times 4^{-\frac23}\right)\frac{f_{X,\max}^{2/3}}{(2p-1)^{1/3}}\frac{1}{L^{\frac{2p-1}{3}}}
	\end{align}
\end{proof}
\section{Effect of Subsampling on the CDF\label{app:Subsampling_CDF}}
Consider the following optimization,
\begin{align}
	\min_{\widehat{f}_{\env}} \quad & \|f-\widehat{f}_{\env}\|_2^2 \nonumber \\
	\text{subject to } \quad & f_{\env}(t) \geq f(t), \quad t \in [0,1].
\end{align}
The same optimization can be re-written in terms of the Fourier Coefficients as,
\begin{align}
	\vec{B}_{\env} := & \argmin_{\vec{B}} \quad \|\vec{A}-\vec{B}\|_2^2 \quad \text{subject to } \nonumber \\
	 & \vec{B}^T \Phi(t) \geq f(t), \quad t \in [0,1], \label{eq:OPT-1}
\end{align}
where $\Phi(t)$ denotes the bandlimited Fourier basis with $(2L+1)$ basis
elements. Suppose, only discrete samples of the signal was available, the
optimization problem to solve would be,
\begin{align}
	\vec{B}_{\app,n}:= & \argmin_{\vec{B}} \quad  \|\vec{A}-\vec{B}\|_2^2 \quad \text{ subject to } \nonumber \\
	 & \vec{B} \Phi(t_n) \geq f(t_n) \quad \text{ for } t_n \in \left\{0,\frac1n, \frac2n, \cdots,1\right\} \label{eq:OPT-2}
\end{align}
Since \eqref{eq:OPT-2} has only a subset of constraints compared to \eqref{eq:OPT-1}, 
\begin{align}
	\|\vec{A}-\vec{B}_{\app,n}\|_2^2 \leq \|\vec{A}-\vec{B}_{\env}\|_2^2.
\end{align}
Using the method in~\cite{Kumar:2017}, we can construct a suboptimal envelope
signal with coefficients $\vec{B}_{\subopt}$ such that $B_{\subopt}[k] =
\begin{cases}B_{\app,n}[k], \quad k \neq 0 \\ B_{\app,n}[0]+\frac{c+c'}{n},
\quad k=0 \end{cases}$, where the constants arise from the assumptions
$|f'(t)|\leq c$ and $f_{\app,n}'(t) \leq c'$, described in
\eqref{eq:discrete_bound}. Since $\vec{B}_{\subopt}$ satisfies the envelope
constraint in \eqref{eq:OPT-1}, we can write 
\begin{align}
	\|\vec{A}-\vec{B}_{\app,n}\|_2^2 \leq \|\vec{A}-\vec{B}_{\env}\|_2^2 \leq \|\vec{A}-\vec{B}_{\subopt}\|_2^2.
\end{align}
Similar to the earlier case, we wish to find an upperbound on CDF difference,
viz. for all $\varepsilon >0$
\begin{align}
	\left|F_X(x)-F_{X_{\app,n}}(x)\right| \leq \varepsilon f_X(x) +
	\frac{1}{\varepsilon^2}\eE\left(\|X-X_{\app,n}\|^2\right). 
\end{align}
Consider, 
\begin{align}
	\|X-X_{\app,n}\|^2 &\leq \|X-X_{\env}\|^2 + \|X_{\env}-X_{\app,n}\|^2
	\quad \text{(Traingle Inequality)} \nonumber \\
	& = \sum_{|k|\leq L}|B_{\env}[k]-A[k]|^2 + \sum_{|k|\leq L} |A[k]|^2 + \sum_{|k|\leq L}|B_{\env}[k]-B_{\app,n}[k]|^2 \nonumber \\
	& \leq 3\sum_{|k|\leq L}|B_{\env}[k]-A[k]|^2 + \sum_{|k|\leq L}|B_{\app,n}[k]-A[k]|^2 \nonumber \\
	& \leq 4\sum_{|k|\leq L}|B_{\subopt}[k]-A[k]|^2 \nonumber \\
	& \leq 4\sum_{|k|\leq L} |B_{\app,n}[k]-A[k]|^2 + 8(B_{\app,n}[0]-A[0])\frac{c+c'}{n} + o(1/n) \nonumber \\
	& \leq \frac{4}{2p-1} \frac{1}{L^{2p-1}} + 8(B_{\app,n}[0]-A[0])\frac{c+c'}{n} + o(1/n)
\end{align}
Thus, 
\begin{align}
	\left|F_X(x)-F_{X_{\app,n}}(x)\right| \leq \varepsilon f_X(x) +
	\frac{1}{\varepsilon^2}\left\{\frac{4}{2p-1} \frac{1}{L^{2p-1}} + 8\mu_{\app,n}\frac{c+c'}{n} + o(1/n)\right\},
\end{align}
where $\mu_{\app,n} =\eE(B_{\app,n}[0]-A[0])$. On assuming $f_X(x)\leq
f_{X,\max}$ and minimizing the upper bound with respect to $\varepsilon$, we get
\begin{align}
	\left|F_X(x)-F_{X_{\app,n}}(x)\right| \leq \left[\left(2^{\frac13} + 2^{-\frac23}\right) f_{X,\max}^{2/3}\right] \left\{\frac{4}{2p-1} \frac{1}{L^{2p-1}} + 8\mu_{\app,n}\frac{c+c'}{n} + o(1/n)\right\}^{1/3}.
\end{align}
{\scriptsize
\bibliography{refs}}

\begin{thebibliography}{10}
\expandafter\ifx\csname url\endcsname\relax
  \def\url#1{\texttt{#1}}\fi
\expandafter\ifx\csname urlprefix\endcsname\relax\def\urlprefix{URL }\fi
\expandafter\ifx\csname href\endcsname\relax
  \def\href#1#2{#2} \def\path#1{#1}\fi

\bibitem{Lalitha:2013}
L.~Sankar, S.~R. Rajagopalan, S.~Mohajer, H.~V. Poor,
  \href{https://doi.org/10.1109/TSG.2012.2211046}{Smart meter privacy: {A}
  theoretical framework}, {IEEE} Transactions on Smart Grid Vol: 4~(2) (2013)
  837--846.
\newblock \href {https://doi.org/10.1109/TSG.2012.2211046}
  {\path{doi:10.1109/TSG.2012.2211046}}.
\newline\urlprefix\url{https://doi.org/10.1109/TSG.2012.2211046}

\bibitem{McMahan:2017}
B.~McMahan, E.~Moore, D.~Ramage, S.~Hampson, B.~A. y~Arcas,
  \href{http://proceedings.mlr.press/v54/mcmahan17a.html}{Communication-efficient
  learning of deep networks from decentralized data}, in: A.~Singh, X.~J. Zhu
  (Eds.), Proceedings of the 20th International Conference on Artificial
  Intelligence and Statistics, Vol.~54 of Proceedings of Machine Learning
  Research, {PMLR}, 2017, pp. 1273--1282.
\newline\urlprefix\url{http://proceedings.mlr.press/v54/mcmahan17a.html}

\bibitem{Konevcny:2016a}
J.~Kone{\v c}n{\'y}, H.~B. McMahan, F.~X. Yu, P.~Richt{\'a}rik, A.~T. Suresh,
  D.~Bacon, \href{https://arxiv.org/abs/1610.05492}{Federated learning:
  Strategies for improving communication efficiency}, in: NIPS Workshop, 2016,
  p.~1.
\newline\urlprefix\url{https://arxiv.org/abs/1610.05492}

\bibitem{Konevcny:2016b}
J.~Kone{\v{c}}n{\'y}, H.~B. McMahan, D.~Ramage, P.~Richt{\'{a}}rik,
  \href{http://arxiv.org/abs/1610.02527}{Federated optimization: Distributed
  machine learning for on-device intelligence}, CoRR abs/1610.02527 (2016) 1.
\newblock \href {http://arxiv.org/abs/1610.02527} {\path{arXiv:1610.02527}}.
\newline\urlprefix\url{http://arxiv.org/abs/1610.02527}

\bibitem{Bonawitz:2019}
K.~Bonawitz, H.~Eichner, W.~Grieskamp, D.~Huba, A.~Ingerman, V.~Ivanov,
  C.~Kiddon, J.~Kone{\v{c}}n{\'y}, S.~Mazzocchi, H.~B. McMahan, T.~V.
  Overveldt, D.~Petrou, D.~Ramage, J.~Roselander,
  \href{http://arxiv.org/abs/1902.01046}{Towards federated learning at scale:
  System design}, CoRR abs/1902.01046 (2019).
\newblock \href {http://arxiv.org/abs/1902.01046} {\path{arXiv:1902.01046}}.
\newline\urlprefix\url{http://arxiv.org/abs/1902.01046}

\bibitem{Kairouz:2021}
P.~Kairouz, H.~B. McMahan, \href{http://dx.doi.org/10.1561/2200000083}{Advances
  and open problems in federated learning}, Foundations and Trends in Machine
  Learning 14~(1) (2021) --.
\newblock \href {https://doi.org/10.1561/2200000083}
  {\path{doi:10.1561/2200000083}}.
\newline\urlprefix\url{http://dx.doi.org/10.1561/2200000083}

\bibitem{Li:2020}
T.~Li, A.~K. Sahu, A.~Talwalkar, V.~Smith,
  \href{https://doi.org/10.1109/MSP.2020.2975749}{Federated learning:
  Challenges, methods, and future directions}, {IEEE} Signal Processing
  Magazine 37~(3) (2020) 50--60.
\newblock \href {https://doi.org/10.1109/MSP.2020.2975749}
  {\path{doi:10.1109/MSP.2020.2975749}}.
\newline\urlprefix\url{https://doi.org/10.1109/MSP.2020.2975749}

\bibitem{Suresh:2017}
A.~T. Suresh, F.~X. Yu, S.~Kumar, H.~B. McMahan,
  \href{http://proceedings.mlr.press/v70/suresh17a.html}{Distributed mean
  estimation with limited communication}, in: Proceedings of the 34th
  International Conference on Machine Learning, {ICML} 2017, Sydney, NSW,
  Australia, 6-11 August 2017, Vol.~70 of Proceedings of Machine Learning
  Research, {PMLR}, 2017, pp. 3329--3337.
\newline\urlprefix\url{http://proceedings.mlr.press/v70/suresh17a.html}

\bibitem{Amiri:2020}
M.~M. Amiri, D.~G{\"{u}}nd{\"{u}}z,
  \href{https://doi.org/10.1109/TWC.2020.2974748}{Federated learning over
  wireless fading channels}, {IEEE} Transactions on Wireless Communications
  19~(5) (2020) 3546--3557.
\newblock \href {https://doi.org/10.1109/TWC.2020.2974748}
  {\path{doi:10.1109/TWC.2020.2974748}}.
\newline\urlprefix\url{https://doi.org/10.1109/TWC.2020.2974748}

\bibitem{AIGoogle:2020}
{Google AI Blog} federated analytics: Collaborative data science without data
  collection,
  \url{https://ai.googleblog.com/2020/05/federated-analytics-collaborative-data.html},
  accessed: 2021-05-19 (2020).

\bibitem{Mammen:2018}
P.~M. Mammen, H.~Kumar, K.~Ramamritham, H.~Rashid,
  \href{https://doi.org/10.1145/3208903.3208941}{Want to reduce energy
  consumption, whom should we call?}, in: H.~Schmeck, V.~Hagenmeyer (Eds.),
  Proceedings of the Ninth International Conference on Future Energy Systems,
  e-Energy 2018, Karlsruhe, Germany, June 12-15, 2018, {ACM}, 2018, pp. 12--20.
\newblock \href {https://doi.org/10.1145/3208903.3208941}
  {\path{doi:10.1145/3208903.3208941}}.
\newline\urlprefix\url{https://doi.org/10.1145/3208903.3208941}

\bibitem{Nilsson:2018}
A.~Nilsson, S.~Smith, G.~Ulm, E.~Gustavsson, M.~Jirstrand,
  \href{https://doi.org/10.1145/3286490.3286559}{A performance evaluation of
  federated learning algorithms}, in: Proceedings of the Second Workshop on
  Distributed Infrastructures for Deep Learning, DIDL@Middleware 2018, Rennes,
  France, December 10, 2018, {ACM}, 2018, pp. 1--8.
\newblock \href {https://doi.org/10.1145/3286490.3286559}
  {\path{doi:10.1145/3286490.3286559}}.
\newline\urlprefix\url{https://doi.org/10.1145/3286490.3286559}

\bibitem{Alistarh:2017}
D.~Alistarh, D.~Grubic, J.~Li, R.~Tomioka, M.~Vojnovic,
  \href{https://proceedings.neurips.cc/paper/2017/hash/6c340f25839e6acdc73414517203f5f0-Abstract.html}{{QSGD:}
  communication-efficient {SGD} via gradient quantization and encoding}, in:
  Advances in Neural Information Processing Systems, 2017, pp. 1709--1720.
\newline\urlprefix\url{https://proceedings.neurips.cc/paper/2017/hash/6c340f25839e6acdc73414517203f5f0-Abstract.html}

\bibitem{Feraudo:2020}
A.~Feraudo, P.~Yadav, V.~Safronov, D.~A. Popescu, R.~Mortier, S.~Wang,
  P.~Bellavista, J.~Crowcroft, Colearn: Enabling federated learning in
  mud-compliant iot edge networks, in: Proceedings of the Third ACM
  International Workshop on Edge Systems, Analytics and Networking, 2020, pp.
  25--30.

\bibitem{Li:2020b}
T.~Li, M.~Sanjabi, A.~Beirami, V.~Smith,
  \href{https://openreview.net/forum?id=ByexElSYDr}{Fair resource allocation in
  federated learning}, in: 8th International Conference on Learning
  Representations, {ICLR} 2020, Addis Ababa, Ethiopia, April 26-30, 2020,
  OpenReview.net, 2020, p.~1.
\newline\urlprefix\url{https://openreview.net/forum?id=ByexElSYDr}

\bibitem{Mohri:2019}
M.~Mohri, G.~Sivek, A.~T. Suresh,
  \href{http://proceedings.mlr.press/v97/mohri19a.html}{Agnostic federated
  learning}, in: K.~Chaudhuri, R.~Salakhutdinov (Eds.), Proceedings of the 36th
  International Conference on Machine Learning, Vol.~97 of Proceedings of
  Machine Learning Research, PMLR, 2019, pp. 4615--4625.
\newline\urlprefix\url{http://proceedings.mlr.press/v97/mohri19a.html}

\bibitem{Smith:2017}
V.~Smith, C.~Chiang, M.~Sanjabi, A.~S. Talwalkar,
  \href{https://proceedings.neurips.cc/paper/2017/hash/6211080fa89981f66b1a0c9d55c61d0f-Abstract.html}{{Federated
  Multi-Task Learning}}, in: Advances in Neural Information Processing Systems,
  2017, pp. 4424--4434.
\newline\urlprefix\url{https://proceedings.neurips.cc/paper/2017/hash/6211080fa89981f66b1a0c9d55c61d0f-Abstract.html}

\bibitem{White:2020}
E.~White, Statistical learning for unimpaired flow prediction in ungauged
  basins, Ph.D. thesis, University of California, Davis (2020).

\bibitem{Stockman:2011}
M.~Stockman, M.~Awad, R.~Khanna, Asymmetrical and lower bounded support vector
  regression for power estimation, in: 2011 International Conference on Energy
  Aware Computing, 2011, pp. 1--6.
\newblock \href {https://doi.org/10.1109/ICEAC.2011.6403624}
  {\path{doi:10.1109/ICEAC.2011.6403624}}.

\bibitem{Dijk:2019}
T.~Van~Dijk, G.~De~Croon, How do neural networks see depth in single images?,
  in: 2019 IEEE/CVF International Conference on Computer Vision (ICCV), 2019,
  pp. 2183--2191.
\newblock \href {https://doi.org/10.1109/ICCV.2019.00227}
  {\path{doi:10.1109/ICCV.2019.00227}}.

\bibitem{Maheshwari:2015}
G.~{Maheshwari}, A.~{Kumar}, Optimal quantization of tv white space regions for
  a broadcast based geolocation database, in: 2016 24th European Signal
  Processing Conference (EUSIPCO), 2016, pp. 418--422.
\newblock \href {https://doi.org/10.1109/EUSIPCO.2016.7760282}
  {\path{doi:10.1109/EUSIPCO.2016.7760282}}.

\bibitem{Tun:2021}
Y.~L. Tun, K.~Thar, C.~M. Thwal, C.~S. Hong, Federated learning based energy
  demand prediction with clustered aggregation, in: 2021 IEEE International
  Conference on Big Data and Smart Computing (BigComp), 2021, pp. 164--167.
\newblock \href {https://doi.org/10.1109/BigComp51126.2021.00039}
  {\path{doi:10.1109/BigComp51126.2021.00039}}.

\bibitem{Konstantinos:2021}
K.~Bountrogiannis, G.~Tzagkarakis, P.~Tsakalides, Data-driven kernel-based
  probabilistic sax for time series dimensionality reduction, in: 2020 28th
  European Signal Processing Conference (EUSIPCO), 2021, pp. 2343--2347.
\newblock \href {https://doi.org/10.23919/Eusipco47968.2020.9287311}
  {\path{doi:10.23919/Eusipco47968.2020.9287311}}.

\bibitem{Saputra:2019}
Y.~M. Saputra, D.~T. Hoang, D.~N. Nguyen, E.~Dutkiewicz, M.~D. Mueck,
  S.~Srikanteswara, Energy demand prediction with federated learning for
  electric vehicle networks, arXiv:1909.00907 (2019).

\bibitem{Anavangot:2020}
V.~Anavangot, A.~Kumar, Algorithms for overpredictive signal analytics in
  federated learning, in: 2020 28th European Signal Processing Conference
  (EUSIPCO), 2021, pp. 1502--1506.
\newblock \href {https://doi.org/10.23919/Eusipco47968.2020.9287390}
  {\path{doi:10.23919/Eusipco47968.2020.9287390}}.

\bibitem{mallat2008}
S.~Mallat, A Wavelet Tour of Signal Processing, Third Edition: The Sparse Way,
  3rd Edition, Academic Press, 2008.

\bibitem{Kumar:2017}
A.~Kumar, Optimal envelope approximation in fourier basis with applications in
  {TV} white space, arXiv:1706.00900 (2017).

\bibitem{Wainwright:2019}
M.~J. Wainwright, High-dimensional statistics: A non-asymptotic viewpoint,
  Vol.~48, Cambridge University Press, 2019.

\bibitem{Bhatia:1993}
R.~Bhatia, Fourier Series, 2nd Edition, Hindustan Book Agency, 1993.

\bibitem{Grimmett:2001}
G.~Grimmett, D.~Stirzaker, {Probability and Random Processes}, Vol.~80, Oxford
  university press, 2001.

\end{thebibliography}
\end{document}